\documentclass[journal]{IEEEtran}

\usepackage{times}
\usepackage{epsfig}
\usepackage{graphicx}
\usepackage{amsmath}
\usepackage{amssymb}
\usepackage{mathtools}
\usepackage{calligra}
\usepackage{color}
\usepackage{array}
\usepackage{hhline}
\usepackage{subfig}
\usepackage{multirow}
\usepackage{caption}
\usepackage{boxhandler}
\usepackage{tabularx}
\usepackage{adjustbox}
\usepackage{verbatim}
\usepackage{bm}
\usepackage{float}
\usepackage{hyperref}

\ifCLASSINFOpdf
\else
\fi

\begin{document}
\title{Iterative Network for Image Super-Resolution}

\author{Yuqing~Liu,
	Shiqi~Wang,~\IEEEmembership{Member,~IEEE,}
	Jian~Zhang,
	Shanshe~Wang,
	Siwei~Ma,~\IEEEmembership{Member,~IEEE,}
	and~Wen~Gao,~\IEEEmembership{Fellow,~IEEE}
	\thanks{Y. Liu is with the School of Software, Dalian University of Technology, 
		Dalian 116620, China (e-mail:liuyuqing@mail.dlut.edu.cn).}
	\thanks{S. Wang is with the Department of Computer Science, City University of Hong Kong, Hong Kong (e-mail: shiqwang@cityu.edu.hk).}
	\thanks{J. Zhang is with the School of Electronic and Computer Engineering, Peking University Shenzhen Graduate School, Shenzhen, China. (e-mail: zhangjian.sz@pku.edu.cn)}
	\thanks{S. Wang, S. Ma, and W. Gao are with the School of Electronics Engineering and Computer Science, Institute of Digital Media, Peking University, Beijing 100871, China (e-mail: sswang@pku.edu.cn; swma@pku.edu.cn; wgao@pku.edu.cn).}
	\thanks{This work was supported by Key-Area Research and Development Program of Guangdong Province (2019B010133001), PKU-Baidu Fund (2019BD003) and High-performance Computing Platform of Peking University, which are gratefully acknowledged.}
}

\markboth{MANUSCRIPT SUBMITTED TO IEEE TMM}%
{Shell \MakeLowercase{\textit{et al.}}: Bare Demo of IEEEtran.cls for IEEE Journals}

\maketitle

\begin{abstract}
	Single image super-resolution (SISR), as a traditional ill-conditioned inverse problem, has been greatly revitalized by the recent development of convolutional neural networks (CNN). These CNN-based methods generally map a low-resolution image to its corresponding high-resolution version with sophisticated network structures and loss functions, showing impressive performances. This paper provides a new insight on conventional SISR algorithm, and proposes a substantially different approach relying on the iterative optimization. A novel iterative super-resolution network (ISRN) is proposed on top of the iterative optimization. We first analyze the observation model of image SR problem, inspiring a feasible solution by mimicking and fusing each iteration in a more general and efficient manner. Considering the drawbacks of batch normalization, we propose a feature normalization (F-Norm, FN) method to regulate the features in network. Furthermore, a novel block with FN is developed to improve the network representation, termed as FNB. Residual-in-residual structure is proposed to form a very deep network, which groups FNBs with a long skip connection for better information delivery and stabling the training phase. Extensive experimental results on testing benchmarks with bicubic (BI) degradation show our ISRN can not only recover more structural information, but also achieve competitive or better PSNR/SSIM results with much fewer parameters compared to other works. Besides BI, we simulate the real-world degradation with blur-downscale (BD) and downscale-noise (DN). ISRN and its extension ISRN+ both achieve better performance than others with BD and DN degradation models.
\end{abstract}

\begin{IEEEkeywords}
	Single image super-resolution, iterative optimization, feature normalization.
\end{IEEEkeywords}

%
\IEEEpeerreviewmaketitle

\section{Introduction}
\IEEEPARstart{S}{ingle} image super resolution~(SISR) is a traditional ill-posed problem in image processing. Given a low-resolution (LR) image, the task of SISR is to find the corresponding image with high-resolution~(HR). Convolutional neural network~(CNN) has shown impressive performance for image restoration~\cite{TMM_Deep1, TMM_Deep2, CSVT_Deep1}. Similar to HDR imaging, it is challenging to preserve high-frequency information for SISR~\cite{[4]}, where fine details act as a critical role~\cite{[5]}. Recently, there are numerous CNN-based works for image super-resolution~\cite{TMM_survey, CSVT_Deep2}. 

As far as we know, SRCNN~\cite{srcnn_dong2015image} is the first CNN-based method for image SR. After SRCNN, researchers build deeper and wider networks with elaborate designs for better performance. RDN~\cite{rdn_zhang2018residual}, RCAN~\cite{rcan_zhang2018image}, SAN~\cite{san_dai2019second} and other recent works achieve state-of-the-art restoration performance with well-designed network structures. These works enjoy a straightforward structure to map LR images to HR images.

Besides the straightforward designs, there are recursive networks for image restoration with restricted parameters. DRCN~\cite{drcn_kim2016deeply}, DRRN~\cite{drrn_tai2017image}, SRFBN~\cite{srfbn_li2019feedback} and other recent works have been considered for effective image SR. Unfortunately, these methods lack an explanation of intrinsic optimization mechanism in nature.

Normalization is also an effective component for boosting the network capacity. Batch normalization (BN) is one of the most important normalization methods for the network design. VDSR~\cite{vdsr_kim2016accurate}, SRResNet~\cite{srgan_ledig2017photo}, and other works utilize BN in the network for image SR. However, BN suffers from texture confusion from mini-batch~\cite{esrgan}, which may be a sub-optimal choice for image SR.

In this paper, an iterative super-resolution network is proposed to solve the SISR problem, termed as ISRN. We analyze the observation model and the target of image SR from the perspective of traditional energy optimization~\cite{ircnn_zhang2017learning, zhang2019deep, gu2019blind}. Motivated by those works, the half quadratic splitting~(HQS) method~\cite{hqs_afonso2010fast} is adopted to analyze the SR problem and obtain a feasible solution. The network is designed based on the solution with iterative structure. Features from each iteration are collected and fused to obtain the final result based on maximum likelihood estimation~(MLE). In vanilla HQS method, degradation model should be given explicitly to find the close-form solution. However, when the degradation models are complex, it is challenging to find a formula description. From this perspective, a network structure is introduced to simulate the degradation and optimization.

\begin{figure}[t]
	\captionsetup[subfloat]{labelformat=empty, justification=centering}
	\begin{center}
		\newcommand{\rowArg}{2cm}
		\newcommand{\fullSize}{4.5cm}
		\newcommand{\fullWidth}{5.7cm}
		\newcommand{\patchSize}{1.85cm}
		\scriptsize
		\setlength\tabcolsep{0.05cm}
		\begin{tabular}[b]{c c c c}
			\multicolumn{3}{c}{\multirow{2}{*}[\rowArg]{
					\subfloat[Example image from Urban100~\cite{Urban100_huang2015single}]
					{\includegraphics[height=\fullSize, width=\fullWidth]
						{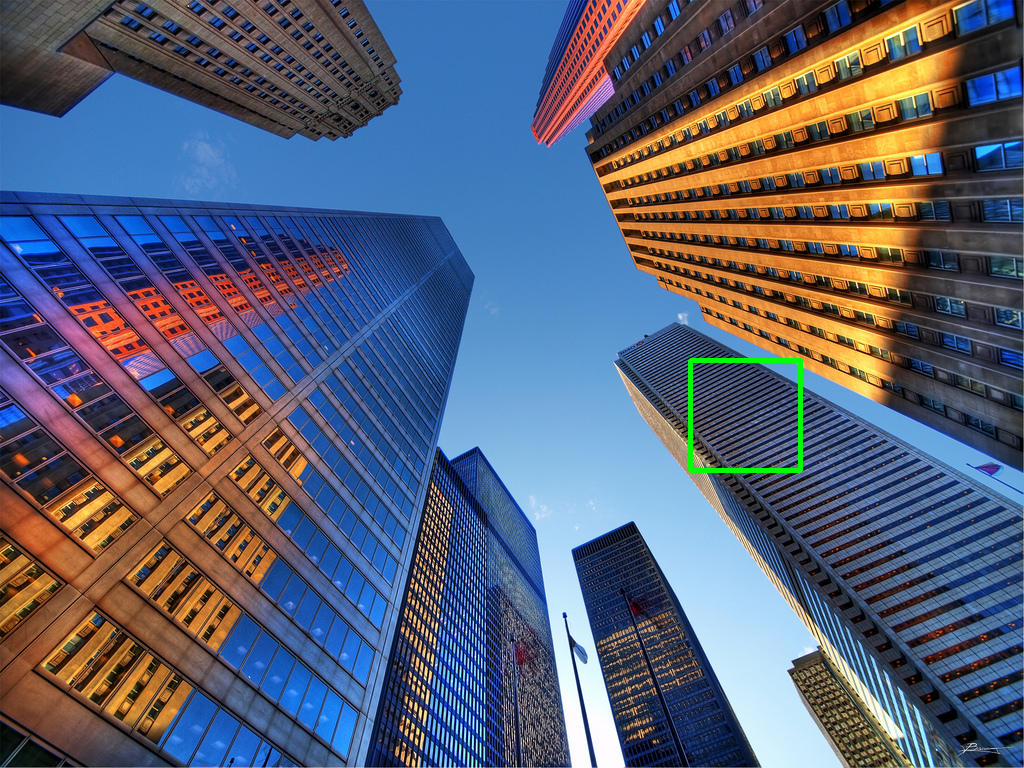}}}} &
			\subfloat[HR~\protect\linebreak(PSNR/SSIM)]
			{\includegraphics[width = \patchSize, height = \patchSize]
				{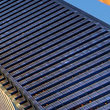}} \\[-0.3cm] & & & 
			\subfloat[Bicubic~\protect\linebreak(22.43/0.6037)]
			{\includegraphics[width = \patchSize, height = \patchSize]
				{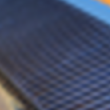}} \\[-0.3cm] 
			\subfloat[RDN~\cite{rdn_zhang2018residual} \protect\linebreak(24.22/0.7552)]
			{\includegraphics[width = \patchSize, height = \patchSize]
				{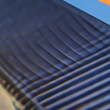}} &
			\subfloat[SAN~\cite{san_dai2019second} \protect\linebreak(24.22/0.7572)]
			{\includegraphics[width = \patchSize, height = \patchSize]
				{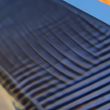}} &
			\subfloat[SRFBN~\cite{srfbn_li2019feedback} \protect\linebreak(24.21/0.7538)]
			{\includegraphics[width = \patchSize, height = \patchSize]
				{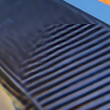}} &
			\subfloat[Ours\protect\linebreak(\textbf{24.34}/\textbf{0.7606})]
			{\includegraphics[width = \patchSize, height = \patchSize]
				{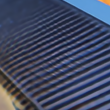}}
		\end{tabular}
	\end{center}
	\setlength{\abovecaptionskip}{0pt plus 2pt minus 2pt}
	\setlength{\belowcaptionskip}{0pt plus 2pt minus 2pt}
	\vspace{-0.2cm}
	\caption{Visual quality comparisons for various image SR methods.}
	\label{Fig:slogan}
\end{figure}

In particular, a novel block with feature normalization~(F-Norm, FN) termed as FNB is designed in ISRN. Different from other normalization methods, the proposed FN learns the weight and bias adaptively, and uses convolutional layer to adjust every pixel of the feature. In this point of view, different pixels will be adjusted in different levels. To pass the features from shallow layers to deeper more efficiently, FNBs are grouped with a residual structure and padding layers, termed as FNG. Extensive experimental results show ISRN and the extension model ISRN$^+$ with self-ensemble are competitive or superior in terms of PSNR/SSIM with much fewer parameters. Subjective visualizations from Fig~\ref{Fig:slogan} clearly show that ISRN can recover structural textures more effectively. Besides bicubic~(\textbf{BI}) degradation, we also simulate the real-world degradation by blur-downscale~(\textbf{BD}) and downscale-noise~(\textbf{DN}) operations. ISRN and ISRN$^+$ perform better on both objective and subjective comparisons with \textbf{BD} and \textbf{DN} degradation models.

The main contributions of this paper are summarized as follows:
\begin{itemize}
	\item We provide a new perspective on SISR by integrating the conventional optimization architecture with deep convolution networks. In this perspective, a novel and lightweight iterative super-resolution network~(ISRN) is proposed.
	
	\item We propose a novel block with feature normalization~(FNB). FNBs are grouped with residual structure and padding layers to bypass the features with skip connections more effectively, termed as FNG.
	
	\item Experimental results show ISRN is competitive or better in terms of PSMR/SSIM with much fewer parameters. Visualization results indicate that ISRN delivers better performance on complex structural information recovery. Furthermore, ISRN and the extension model ISRN$^+$ can achieve better performance in terms of both subjective and objective comparisons with \textbf{BD} and \textbf{DN} degradation models.
\end{itemize}

\section{Literature Review}
SRCNN~\cite{srcnn_dong2015image} proposed by Dong~\textit{et al.} is the first work for SISR problem with a three-layer network, which achieves better performance than traditional methods. The three layers of SRCNN are corresponding to the steps of traditional sparse coding methods. Deeper networks usually result in better performance. Kim~\textit{et al.} increased the layer number and introduced global residual learning in VDSR~\cite{vdsr_kim2016accurate} for stronger network representation and better performance. Deconvolution layer was widely used in early SISR works for resolution increase. ESPCN~\cite{espcn_shi2016real} proposed by Shi~\textit{et al.} substituted the deconvolution with sub-pixel convolutional layer for more effective upscaling operation, which has been proved to an effective structure. After ESPCN, most of the SISR works choose sub-pixel layer instead of deconvolution. Residual structure has shown amazing performance on image and video restoration~\cite{san_dai2019second}. To obtain better performance, in EDSR~\cite{edsr_lim2017enhanced} proposed by Lim~\textit{et al.}, the residual blocks with more filters have been adopted. Batch normalization layers in EDSR are removed to decrease the memory cost and build the network deeper. Besides deeper designs, there are works concentrating on effective blocks. Since dense connection has shown good performance for different tasks, SRDenseNet~\cite{srdensenet_tong2017image} proposed by Tong~\textit{et al.} stacked dense blocks for better performance. Zhang~\textit{et al.} combined residual and dense connections in RDN~\cite{rdn_zhang2018residual}. He~\textit{et al.} designed ODENet~\cite{odenet_he2019ode} based on ordinary differential equations. As one of the pioneering works adopting residual-in-residual (RIR) structure, ESRGAN~\cite{esrgan} achieves good visualization restoration performance. Multi-scale designs also turn out to be an effective component~\cite{CSVT_Multi, TMM_MRFN}. MRFN~\cite{TMM_MRFN} introduced a multi-receptive-field design for feature exploration. Meanwhile, there are also works focusing on the attention mechanism~\cite{CSVT_Attention, CSVT_Deep2}.

Recursive designs have also been widely studied for image restoration problems. To our best knowledge, Kim~\textit{et al.} firstly applied recursive structure with share convolution layers in DRCN~\cite{drcn_kim2016deeply} for SISR problem. To expand the receptive fields, DRCN increased the network depth by using shared filters with limited parameters. Inspired by the residual design, Tai~\textit{et al.} proposed DRRN~\cite{drrn_tai2017image} with residual blocks incorporated. DRRN introduced a recursive block design with the combination of convolution layers, achieving better performance than VDSR. MemNet~\cite{memnet_tai2017memnet} developed by Tai~\textit{et al.} is motivated by long-term memory model of human's brain. In MemNet, recursive and gate units are proposed to simulate the memory mechanism, and memory blocks have been adopted for better performance. Recently, Yang~\textit{et al.} designed DRFN~\cite{TMM_DRFN} with recurrent structure for large factors. However, these methods lack an explanation of intrinsic optimization mechanism in nature.

There are numerous normalization methods developed for network representation improvement. VDSR used batch normalization~(BN)~\cite{ioffe2015batch} between different convolution layers. Since BN consumes more memory~\cite{survey_wang2019deep}, recent works have replaced the normalization with more efficient convolutional layers. Weight normalization~(WN) was adopted in WDSR~\cite{wdsr_yu2018wide} proposed by Yu~\textit{et al.}, which was firstly proposed by Salimans~\textit{et al.} for recurrent models~\cite{weightnorm}.

\section{Formulaic Analysis for Image Super-Resolution}

The observation model of SISR problem could be formulated as,
\begin{equation}
	\label{Eq:1}
	\mathbf{I}^{LR}=\mathcal{D}(\mathbf{I}^{HR})+\mathbf{n},
\end{equation}
where $\mathcal{D}(\cdot)$ is the degradation operator, $\mathbf{n}$ is the noise term, and $\mathbf{I}^{LR}, \mathbf{I}^{HR}$ are LR and HR images respectively. Generally speaking, $\mathcal{D}(\cdot)$ could be a bicubic down-sampler, blur kernel or the mixture operations. 

Given an LR image $\mathbf{I}^{LR}$, the target of super-resolution is to find an $\mathbf{I}^{SR}$ satisfying,
\begin{equation}
	\label{Eq:2}
	\mathbf{I}^{SR}=\arg\min_{\mathbf{I}^{SR}}\frac{1}{2}||\mathcal{D}(\mathbf{I}^{SR})-\mathbf{I}^{LR}||^2_\ell+\lambda\phi(\mathbf{I}^{SR}),
\end{equation}
where $\phi(\cdot)$ is the image prior term and $\lambda$ is a factor. $||\cdot||_\ell$ means the $\ell$-norm.

To obtain the HR image, there are numerous CNN-based works calculating a direct mapping from LR to HR, aiming to solve Eqn.~(\ref{Eq:2}). In this paper, half quadratic splitting~(HQS) \cite{hqs_afonso2010fast, hqs_geman1992constrained} method is applied for finding the solutions. Let $\mathbf{u}=\mathcal{D}(\mathbf{I}^{SR})$, then Eqn.~(\ref{Eq:2}) could be re-written as,
\begin{equation}
	\label{Eq:3}
	\begin{split}
		\mathbf{I}^{SR}=\arg\min_{\mathbf{I}^{SR}}\frac{1}{2}||\mathbf{u}-\mathbf{I}^{LR}||^2_\ell+\lambda\phi(\mathbf{I}^{SR}), \\
		\text{s.t. } \mathbf{u}=\mathcal{D}(\mathbf{I}^{SR}).
	\end{split}
\end{equation}

As such, Eqn.~(\ref{Eq:3}) could be solved in an iterative way by calculating $\mathbf{I}^{SR}_{k}$ and $\mathbf{u}_{k}$ alternatively, 
\begin{equation}
	\label{Eq:4}
	\mathbf{I}^{SR}_{k}=\arg\min_{\mathbf{I}^{SR}}\frac{\beta_k}{2}||\mathcal{D}(\mathbf{I}^{SR})-\mathbf{u}_{k-1}||^2_\ell+\lambda\phi(\mathbf{I}^{SR}),
\end{equation}
\begin{equation}
	\label{Eq:5}
	\mathbf{u}_{k}=\arg\min_\mathbf{u}\frac{1}{2}||\mathbf{u}-\mathbf{I}^{LR}||^2_\ell+\frac{\beta_k}{2}||\mathbf{u}-\mathcal{D}(\mathbf{I}^{SR}_{k})||_\ell^2,
\end{equation}
where $\beta_k$ is a weighting factor for the $k$-th iteration and varies in a non-descending order for each iteration.
For Eqn.~(\ref{Eq:5}), $\mathbf{u}$ has the closed-form solution by linearly combining $\mathbf{I}^{LR}$ and $\mathcal{D}(\mathbf{I}^{SR})$. 

Let $\phi_{\beta_k}=\frac{1}{\beta_k}\phi$, then Eqn.~(\ref{Eq:4}) can be re-written as:
\begin{equation}
	\label{Eq:6}
	\mathbf{I}^{SR}_{k}=\arg\min_{\mathbf{I}^{SR}}\frac{1}{2}||\mathcal{D}(\mathbf{I}^{SR})-\mathbf{u}_{k-1}||^2_\ell+\lambda\phi_{\beta_k}(\mathbf{I}^{SR}).
\end{equation}

\begin{figure}[t]
	\centering
	\includegraphics[width=\linewidth]{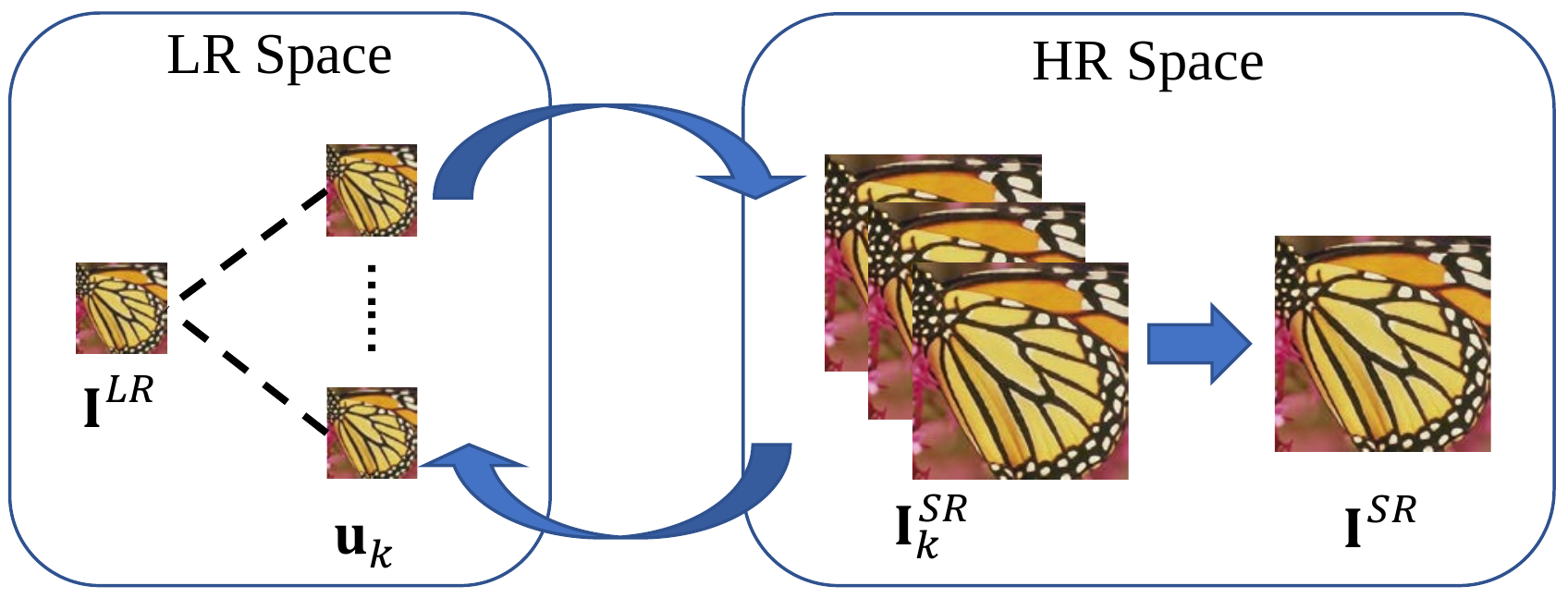}
	\caption{A simple illustration of our proposed iterative scheme. The distance shrinks on LR space for each iteration, and the results are optimized on HR space.}
	\label{Fig:iteration}
\end{figure}

The iterative solution can be interpreted from another perspective. In particular, Eqn.~(\ref{Eq:6}) can be cast into a mapping from the LR space to HR space, such that a reasonably good result on average can be obtained in each iteration. Eqn.~(\ref{Eq:5}) aims to achieve a linear combination of $\mathbf{I}^{LR}$ and $\mathcal{D}(\mathbf{I}^{SR})$, which can be regarded as guiding $\mathbf{I}^{LR}$ with a specific direction $(\mathbf{I}^{LR}-\mathcal{D}(\mathbf{I}^{SR}))$ and a specific step length governed by the parameter $\beta_k$. This is in analogous to the gradient descent method. Since the exact distance between $\mathbf{I}^{HR}$ and $\mathbf{I}^{SR}$ on HR space is unknown, these iterative steps shrink the distance between $\mathbf{I}^{LR}$ and $\mathcal{D}(\mathbf{I}^{SR})$. As such, we hold the notion that iterative optimization steps gradually decrease the distance on the HR space by adjusting the distance on the LR space. An illustration of the steps could be demonstrated in Fig.~\ref{Fig:iteration}.

However, there are two critical issues. On one hand, the down-sampler operator $\mathcal{D}(\cdot)$ which accounts for the mapping from the HR space to LR space is difficult to be simulated. In general, $\mathcal{D}(\cdot)$ could be regarded as a bicubic down-sampling operator while training. However, in some complicated situations, it could be difficult to explicitly express $\mathcal{D}(\cdot)$. From Eqn.~(\ref{Eq:5}), the accuracy of $\mathcal{D}(\cdot)$ directly influences the optimization. From this perspective, the degradation model should be learned from paired data. On the other hand, the solution of Eqn.~(\ref{Eq:5}), i.e. $\mathbf{u}_{k}$, is a linear combination of $\mathbf{I}^{LR}$ and $\mathcal{D}(\mathbf{I}^{SR}_{k})$ on $k$-th step, which is close to the one-step gradient descent operation. When the $(k+1)$-th iteration begins, the start point is still $\mathbf{I}^{LR}$ instead of $\mathbf{u}_{k}$. This shows the optimization is memory-less. In other words, the history descent directions do not influence the starting point but only the next descent direction. To handle this issue, outputs from different iterations should be collected and considered jointly to find the final result. It can be regarded as a maximum likelihood estimation~(MLE), demonstrated as,
\begin{equation}
\label{Eq:7}
	\mathbf{I}^{SR}=\arg\max_\mathbf{I}P(\mathbf{I} |\{\mathbf{I}^{SR}_i\}_{i=1}^{K}),
\end{equation}
where $\mathbf{I}^{SR}$ denotes the final HR image, and $\mathbf{I}^{SR}_i$ denotes the output of $i$-th iteration.

Iterative super-resolution network~(ISRN) is designed based on the previous formulation study. From the problem formulation, $\mathbf{I}^{SR}_{k}$ for $k$-th iteration is optimized from Eqn.~(\ref{Eq:6}), which could be cast into an independent super-resolution problem mapping the input LR image $\mathbf{u}_{k-1}$ to HR image $\mathbf{I}^{SR}_{k}$. From this observation, a solver for image super-resolution is suitable to find the solution. We design a network module to find the result, termed as \textit{Solver SR}. While training, the implicit expression of $\mathcal{D}(\cdot)$ and $\phi(\cdot)$ will be learned from the paired data, and the adaptive optimization will be performed. \textit{Solver SR} is shared for each iteration to find the suitable mapping relations between LR space and HR space while training.

The closed-form solution of Eqn.~(\ref{Eq:5}) is a linear combination of $\mathbf{I}^{LR}$ and $\mathcal{D}(\mathbf{I}_{k}^{SR})$. Since there is no explicit expression for $\mathcal{D}(\cdot)$ when degradation models are complex, it is hard to find $\mathcal{D}(\mathbf{I}_{k}^{SR})$ while given $\mathbf{I}_{k}^{SR}$. We investigate a network module to simulate the degradation, and term it as \textit{Down-sampler}. Furthermore, considering that the weighting factor $\beta$ in Eqn.~(\ref{Eq:5}) varies in different iterations, we utilize a network module to learn feasible factors for each iteration and find the solution, which is term as \textit{Solver LR}.

From the formulation, the optimization steps for each iteration are memory-less. It is necessary to collect the outputs of different iterations, and find a suitable result considering all descent directions.  The MLE step could be designed as a network module to find the $\mathbf{I}^{SR}$ with maximum probability, termed as \textit{Solver MLE}.

The comparison between our CNN-based structure and the HQS method is shown in Table~\ref{Tab:hqs}. For every column, we aim to solve the equation by the corresponding CNN-based structure. We compare the components in ISRN and steps in HQS, which shows the relationship between our network and optimization steps.

\begin{figure*}[t]
	\centering
	\includegraphics[width=0.8\linewidth]{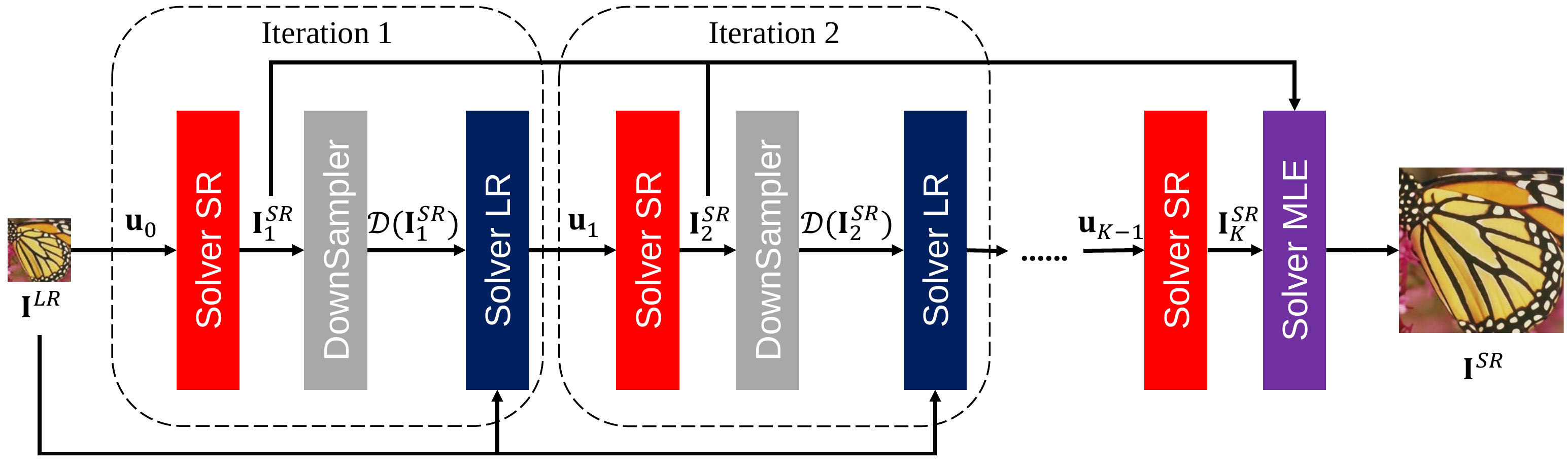}
	\caption{The network structure of the proposed iterative super-resolution network~(ISRN). There are four components in ISRN: \textit{Solver SR}, \textit{Solver LR}, \textit{Down-Sampler}, and \textit{Solver MLE}, corresponding to different steps in formulation study. \textit{Solver SR} is shared for each iteration to find the suitable mapping from LR space to HR space.}
	\label{Fig:network-structure}
\end{figure*}

\begin{table}[t]
    \centering
    \caption{The comparison between ISRN and HQS.}
    \label{Tab:hqs}
    \begin{tabular}{|c|c|c|c|c|}
    \hline
         \textbf{ISRN}& \textit{Solver SR}& \textit{Solver LR}& \textit{Downsampler}& \textit{Solver MLE} \\
         \hline
         \textbf{HQS}& Eqn.~(\ref{Eq:6})&  Eqn.~(\ref{Eq:5})& $\mathcal{D}(\cdot)$&  Eqn.~(\ref{Eq:7}) \\
    \hline
    \end{tabular}
\end{table}

\section{Network Design}
As shown in Fig.~\ref{Fig:network-structure}, there are four modules in the proposed ISRN, corresponding to \textit{Solver SR}, \textit{Solver LR}, \textit{Down-sampler} and \textit{Solver MLE} separately. Herein, these modules are detailed as follows.

\textbf{\textit{Solver SR}} is the main component to generate images in HR space from the LR space shared for every iteration, which is formulated as $\mathcal{SR}(\cdot)$. Most recent networks for image restoration are deep, which may accumulate the feature variance. Batch normalization is proposed for performance improvement, which may consume much memory~\cite{survey_wang2019deep}. In this paper, a novel feature normalization~(F-Norm) method is proposed, formulated as,
\begin{equation}
	f^o_m = (g_m * f^i_m + b_m) + f^i_m,
\end{equation}
where $m$ is the channel index, $f^i$ and $f^o$ are corresponding input and output feature channels, $g$ is a convolution kernel, and $b$ is the bias. To preserve the original feature information, the features before and after normalization are added as the final output.

The proposed FN is designed with the hypothesis that different channels contain different information. Different channels are treated parallelly to prevent the information fusion. The parallel normalization will decrease the parameters and computation complexity, making it flexible for various network designs.

The FN has a similar formulation with BN. If $g_k$ is regarded as a convolution kernel with size $1\times1$, then it holds a same operation with BN when setting batch size as 1. The F-Norm performs normalization on features independently, preventing the influence of minibatch in BN. The factors for normalization are explored from the only feature maps. Different form BN, F-Norm is implemented with only one convolution layer, which is fast and with little memory cost.

Termed as a normalization method, FN derives from BN and has been modified for SISR problem. Firstly, Gaussian distribution in BN is not suitable for SISR problem which gets rid of range flexibility from networks~\cite{edsr_lim2017enhanced}. The distribution adjustment in BN requires large memory cost, making it hard to build a deeper network~\cite{survey_wang2019deep}. FN removes the distribution adjustment. Secondly, the mini-batch in BN may confuse the diverse textures and lead to a bad result~\cite{esrgan}. Different from learning the batch-wise normalization factors, FN adjusts the feature maps in the spatial-wise perspective, which treats every pixel differently. To extremely avoid the confusion, each channel is processed independently by group convolution. Finally, to keep the gradient transmission efficiency as BN, FN adds a residual.

A novel block named feature normalization block~(FNB) is proposed with FN. In FNB, FN is applied at the bottom of residual block. On one hand, it could normalize the feature maps after non-linear processing. On the other hand, using only one normalization layer in each block could save the parameters and computation cost.

Residual structure can gradually pass the shallow layer features to deeper layers. To speed up the feature delivery and make better use of shallow layer features, residual-in-residual~(RIR) structure is applied in the network. A group of FNBs with a skip connection is proposed, termed as FNG. For each FNB in the group, there is a residual structure. The global FNG also acquires a shortcut to pass the shallow features to the deeper and improve the gradient transmission.

There is a padding structure after FNBs, composed of two convolution layers with ReLU activation and a F-Norm layer. This padding structure could introduce a non-linear processing step for main path information. In FNG, F-Norm layer following the last convolution layer aims to normalize the features on the main path. 

\begin{figure*}[t]
	\centering
	\includegraphics[width=0.8\linewidth]{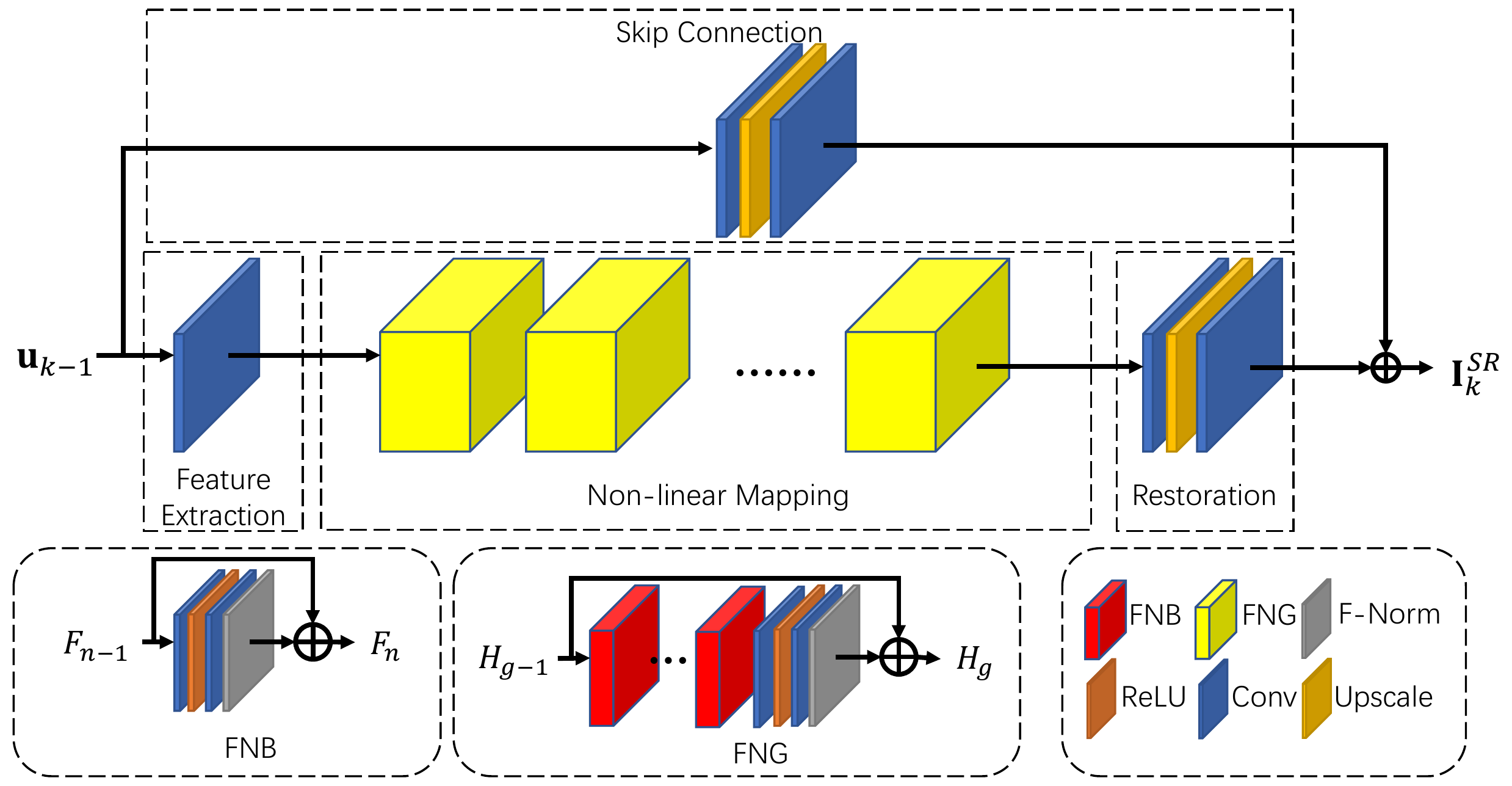}
	\caption{Framework of the \textit{Solver SR} and its components. There are four modules in \textit{Solver SR}, mapping the features from LR space onto HR space.}
	\label{Fig:structure-of-solver-sr}
\end{figure*}

The entire network structure of \textit{Solver SR} is shown in Fig.~\ref{Fig:structure-of-solver-sr}. In analogous to other super-resolution networks, \textit{Solver SR} has a main enhancement path and a skip bypass, which form the global residual framework. The bypath in \textit{Solver SR} upscales the $x_{in}$ by convolution and sub-pixel layer. A convolution layer is applied after each sub-pixel layer to introduce the spatial correlation.

\textit{Solver SR} could be regarded as an complete network structure for single image super-resolution, since it directly maps LR image into HR space. There are four modules in the \textit{Solver SR}. The first convolution layer in the main path denotes the feature extraction module. After feature extraction, several FNGs are used to compose the non-linear mapping module. The restoration module is made up of two convolution layers with a sub-pixel layer. Finally, a skip connection is applied as the shortcut.

Different from RCAN~\cite{rcan_zhang2018image} and other RIR-based works, there is no global residual connection in non-linear mapping module. On one hand, there is a residual structure in proposed FNG. With the stack of FNGs, information could be fully delivered on the shortcuts from top to bottom. On the other hand, the skip module could be regarded as a global residual connection of the entire network, helping the information transmission.

\textbf{\textit{Solver LR}} is a network proposed to solve the Eqn.~(\ref{Eq:5}), formulated as $\mathcal{LR}(\cdot)$. Although Eqn.~(\ref{Eq:5}) has a closed-form solution, the result $\mathbf{u}_{k}$ is a linear combination of $\mathbf{I}^{LR}$ and $\mathcal{D}(\mathbf{I}_{k}^{SR})$, which implies the SR solution will fall into a space spanned by $\mathcal{SR}(\mathbf{I}^{LR})$ and $\mathcal{SR}(\mathcal{D}(\mathbf{I}_{k}^{SR}))$.  Meanwhile, the weight factor $\beta_k$ varies for every iteration. From this persepctive, a network is designed which both introduces the non-linearity and adaptive factors. \textit{Solver LR} aims to adaptively learn the weighting factor and performs the optimization based on the network representation capacity.

The structure of \textit{Solver LR} is a 3-layer network with ReLU activation after the second convolution layer. The first convolution layer aims to linearly combine the feature of $\mathbf{I}^{LR}$ and $\mathbf{u}$. The second layer with ReLU activation introduces the non-linearity. The last convolution layer maintains the same channel number of inputs and the output.

\textit{Solver LR} has a similar structure to SRCNN~\cite{srcnn_dong2015image}, which has been proved effective for filtering. Different form the closed-form solution of Eqn.~(\ref{Eq:5}) which could be regarded as a point-wise operation, the filter-based method \textit{Solver LR} enlarges the receptive field to consider the information nearby.  

\begin{figure}[t]
	\centering
	\includegraphics[width=0.8\linewidth]{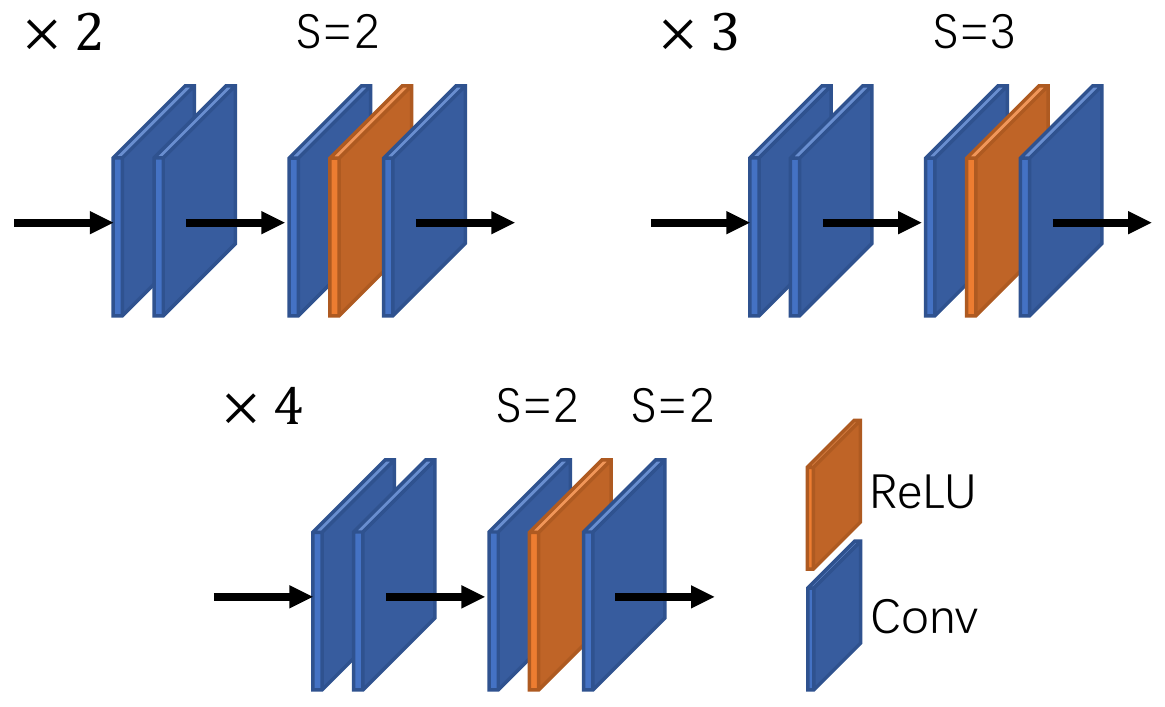}
	\caption{Illustration of the \textit{Down-sampler} with different scaling factors.}
	\label{Fig:structure-of-down-sampler}
\end{figure}

\begin{table*}[t]
    \centering
    \caption{Parameters and MACs comparison with $\textbf{BI}\times4$ degradation.}
    \label{Tab:flops}
    \begin{tabular}{|c||c|c||c|c|c|c|c|}
    \hline
    \textbf{Model}& \textbf{MACs (G)}& \textbf{Params (M)}& \textbf{Set5}& \textbf{Set14}& \textbf{B100}& \textbf{Urban100}& \textbf{Manga109} \\
    \hline
    \hline
    ISRN& 988.8& 3.45& 32.55/0.8992& 28.79/0.7872& 27.74/0.7422& 26.64/0.8033& 31.16/0.9166 \\
    D-DBPN~\cite{dbpn_haris2018deep}& 5213.0& 10.42& 32.47/0.8980& 28.82/0.7860& 27.72/0.7400& 26.38/0.7946& 30.91/0.9137\\
    SRFBN~\cite{srfbn_li2019feedback}& 7466.1& 3.63& 32.47/0.8983& 28.81/0.7868& 27.72/0.7409& 26.60/0.8015& 31.15/0.9160\\
    EDSR~\cite{edsr_lim2017enhanced}& 2895.8& 43.08& 32.46/0.8968& 28.80/0.7876& 27.71/0.7420& 26.64/0.8033& 31.02/0.9148\\
    RCAN~\cite{rcan_zhang2018image}& 919.1& 17.14& 32.63/0.9002& 28.87/0.7889& 27.77/0.7436& 26.82/0.8087& 31.22/0.9173\\
    \hline
    ISRN (K=1)& 187.4& 3.20& 32.30/0.8965& 28.69/0.7845& 27.64/0.7388& 26.30/0.7938& 30.72/0.9116 \\
    MSRN~\cite{msrn_li2018multi}& 368.6& 6.37& 32.26/0.8960& 28.63/0.7836& 27.61/0.7380& 26.22/0.7911& 30.57/0.9103 \\
    OISR-RK2~\cite{oisr}& 412.2& 5.50& 32.32/0.8965& 28.72/0.7843& 27.66/0.7390& 26.37/0.7953& - \\
    CARN~\cite{carn_ahn2018fast}& 90.9& 1.59& 32.13/0.8937& 28.60/0.7806& 27.58/0.7349& 26.07/0.7837& 30.40/0.9082 \\
    \hline
    \end{tabular}
\end{table*}

\begin{figure*}[t]
    \begin{center}
        
    \begin{tabular}[b]{c}
    \subfloat[]{\includegraphics[width=0.8\linewidth]{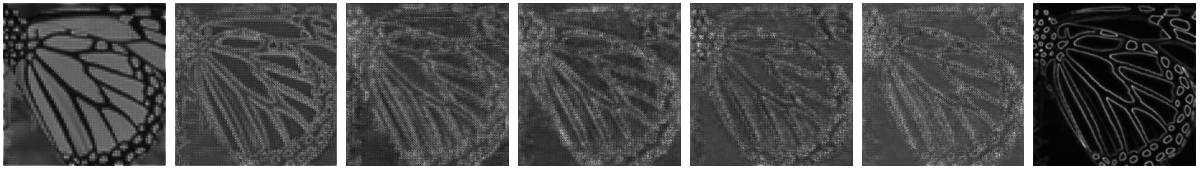}} \\
    \subfloat[]{\includegraphics[width=0.8\linewidth]{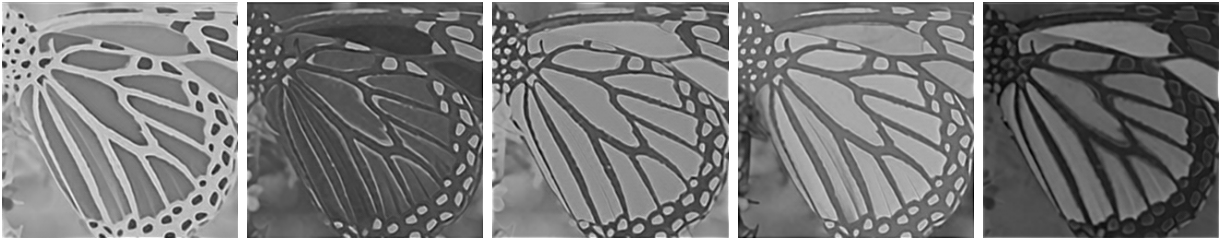}}
    \end{tabular}
    
    \end{center}
    \caption{Visualization comparisons from different iterations. From left to the right, the iteration number increases progressively. Each feature has been enhanced using the same grayscale colormap for visibility.}
    \label{Fig:dbpn-diff}
\end{figure*}

\textbf{\textit{Down-sampler}} is a network dedicated to simulate $\mathcal{D}(\cdot)$. In previous works, the degradation model is usually chosen as bicubic-down, which has an explicit formulation. However, when the degradation is more general or even unknown, it is difficult to calculate $\mathcal{D}(\mathbf{I}^{SR})$. To address this issue, a network is designed to simulate the degradation while training. \textit{Down-sampler} is designed with 4 convolution layers. Considering the mechanism of vanilla bicubic-down, where one pixel corresponds to a $4\times 4$ window while interpolation, the first 2 layers extract the features with the kernel size as 3, which equals a $5\times 5$ receptive field. To simulate the down-sampling operation, two convolution layers with different strides are applied at the subsequence with ReLU activation. Notice that the stride size should be no larger than the kernel size to prevent the information loss. When the scaling factors are $\times2$ and $\times3$, the strides are performed on the first layer. When the scaling factor is $\times4$, the strides are performed as $\times2$ and $\times2$ on both two layers. The structure of \textit{Down-sampler} is shown in Fig.~\ref{Fig:structure-of-down-sampler}.

\textbf{\textit{Solver MLE}} is designed to simulate the maximum likelihood estimation, formulated as $\mathcal{MLE(\cdot)}$. \textit{Solver MLE} is used to analyze $\mathbf{I}^{SR}_k$ from every step and estimate a final result $\mathbf{I}^{SR}$. This model is designed as 2 convolution layers' network with a ReLU activation.

\textbf{Processing step} can be demonstrated as follows. Given an LR input $\mathbf{I}^{LR}$, the input of the first iteration is $\mathbf{u}_0=\mathbf{I}^{LR}$. For the $k$-th step, there is
\begin{equation}
	\mathbf{I}^{SR}_{k} = \mathcal{SR}(\mathbf{u}_{k-1}),
\end{equation}
and the input of $(k+1)$-th iteration is:
\begin{equation}
	\mathbf{u}_{k}=\mathcal{LR}_{k}([\mathcal{D}_{k}(\mathbf{I}^{SR}_{k}), \mathbf{I}^{LR}]).
\end{equation}

\textit{Solver SR} is shared for every iteration, while \textit{Solver LR} and \textit{Down-sampler} are different. We hold the notion that the difference of \textit{Solver LR} and \textit{Down-sampler} could be diverse in terms of the input space and finally enhance the final result. The output of the network is given by,
\begin{equation}
	\mathbf{I}^{SR} = \mathcal{MLE}([\mathbf{I}^{SR}_1,...,\mathbf{I}^{SR}_K]).
\end{equation}

\section{Discussion}
\textbf{Comparisons with RCAN}~\cite{rcan_zhang2018image}. In RCAN, residual-in-residual is embedded with Squeeze-and-Excitation~\cite{senet_hu2018squeeze} block to perform channel attention. Different from RCAN, in ISRN, an iterative structure is designed for better performance with fewer parameters. At the same time, ISNR concentrates on feature normalization rather than channel attention. ISRN applies the feature normalization method, and useful evidences have been provided. RCAN aims to find direct mapping from LR space to HR space, while ISRN provides an optimization perspective for finding solution. By devising these elaborate designs, ISRN achieves competitive performance with \textbf{BI}$\times4$ degradation with much fewer parameters and near computation complexity than RCAN, as shown in Table~\ref{Tab:flops}.

\textbf{Comparisons with SRFBN}~\cite{srfbn_li2019feedback}. SRFBN applies a feedback mechanism to recursively enhance the super-resolution performance, which directly concatenates shallow and deep features. Different from SRFBN, ISRN provides a mathematical proof of the model, and simulate each solver with corresponding network components. ISRN feeds the network with different inputs in every iteration. SRFBN is trained with outputs from every iterations, while ISRN is trained with only one output from \textit{Solver MLE}.

\textbf{Comparisons with IRCNN}~\cite{ircnn_zhang2017learning}. There are two ways for converting the Eqn.~(\ref{Eq:3}), by using $\textbf{u}=\textbf{I}^{SR}$ or $\textbf{u}=\mathcal{D}(\textbf{I}^{SR})$. If we choose the first one, then it is same with ICRNN, where Eqn.~(\ref{Eq:6}) will be converted into a denoising sub-problem. It leads to high computational complexity with the increasing of feature resolution. Some details may also be removed by the denoiser network. From this point of view, ISRN achieves better performance than IRCNN. Besides the different splitting method, ISRN is regarded as an end-to-end network, instead of building the pipeline as plug-and-play.

\textbf{Comparisons with DBPN}~\cite{dbpn_haris2018deep}. DBPN applies a back-projection method for iterative up-and-down sampling and concerns the residuals in two space. ISRN provides a different perspective for SISR problem, and builds the pipeline based on mathematical analysis. There are two main differences between ISRN and DBPN. Firstly, DBPN adjusts the distance in LR and HR spaces jointly by up- and down-projection blocks. In contrast, ISRN shrinks the distance only in HR space. As addressed in the formulation study, we do not exactly know the distance between $\textbf{I}^{SR}$ and $\textbf{I}^{HR}$ in HR space. HQS optimizes the LR results by performing gradient descent-like step, and provides a reliable descent direction for restoration. The optimization in LR space will decrease the computation complexity because of the smaller resolution. As such, ISRN achieves better restoration performance than DBPN with lower computational complexity. Table~\ref{Tab:flops} shows the computation complexity and performance comparison with $\textbf{BI}\times4$ degradation.

Secondly, in DBPN, the projected feature for each iteration is a concatenation of all previous processed ones. The dense connection aims to build an efficient gradient transmission pathway. However, the diverse inputs confuse the descent direction. In other words, as mentioned in our paper, each iteration provides a different direction for shrinking the distance. Dense connection in DBPN makes the iteration mechanism inefficient. ISRN provides an intrinsic different view for the input of each iteration. In ISRN, $\textbf{I}^{LR}$ acts an anchor to find a specific direction for gradient descent, and guarantees the efficiency of iteration mechanism. Fig.~\ref{Fig:dbpn-diff} (a) and (b) demonstrate the DBPN and ISRN outputs in HR space separately. Each feature has been enhanced using the same grayscale colormap for visibility. From the visualization comparison, each iteration in ISRN learns a specific different direction between $\mathbf{I}^{SR}$ and $\mathbf{I}^{HR}$. In contrary, DBPN learns a comprehensive descent direction and makes the output feature map more unnatural with artifacts.

\textbf{Plug-and-Play}. Since \textit{Solver SR} is an independent network for SR, it is feasible to consider building the pipeline as plug-and-play. We hold the hypothesis that a straightforward image restoration network can be regarded as a sparse-coding like solver. After trained with $\ell_1$ loss, the network will find a best mapping on average. Notice that different networks learn different coding dictionaries, which vary widely. It is difficult to fit general parameters for other components. From this perspective, the proposed ISRN is regarded as an end-to-end structure rather than plug-and-play.

\section{Experimental Results}
\subsection{Settings}
In ISRN, all layers are with kernel size as $3\times 3$ except for the skip bypath in \textit{Solver SR} and all layers after sub-pixel. These layers are with kernel size as $5\times 5$ for a larger receptive. Layers of \textit{Solver SR}, \textit{Solver MLE}, and \textit{Solver LR} have $M=64$ filters, and layers of \textit{Down-sampler} have $M_0=32$ filters. For each FNP, there are $N=6$ FNBs; and for \textit{Solver SR}, there are $G=6$ FNGs. There are $K=5$ iterations in the network. During training, the $\ell_1$ loss is chosen as loss function.

In the training progress, 800 images are used from DIV2K~\cite{div2k_timofte2017ntire} dataset for training, and 5 images are used for validation. Five benchmark datasets are used for testing: Set5~\cite{set5_bevilacqua2012low}, Set14~\cite{set14_zeyde2010single}, B100~\cite{B100_martin2001database}, Urban100~\cite{Urban100_huang2015single} and Manga109~\cite{Manga109_matsui2017sketch}. Images from B100 are from real-world containing rich high-frequency information. There are numerous buildings in Urban100, such that abundant straight textures are included. Manga109 are cartoons with structural information. Three benchmark degradation models are used to simulate LR images: bicubic~(\textbf{BI}) , blur-downscale~(\textbf{BD}), and downscale-noise~(\textbf{DN}). All the parameter settings of degradation models are identical with RDN~\cite{rdn_zhang2018residual}. Adam optimizer, which is widely used in several super-resolution tasks~\cite{rdn_zhang2018residual, rcan_zhang2018image, san_dai2019second}, is used with learning rate $lr=10^{-4}$. The learning rate is halved for every 200 epochs. The patch size of LR inputs is $48\times 48$. The training data are augmented by randomly flipping and rotation. In total, the network is trained with 1000 iterations. Self-ensemble~\cite{rdn_zhang2018residual} is adopted to improve the performance of ISRN, and the extended model is named as ISRN$^+$. The source code and pre-trained models of ISRN and ISRN$^+$ can be downloaded at: \url{https://github.com/yuqing-liu-dut/ISRN}.

\begin{table*}[t]
	\centering
	\caption{Average PSNR/SSIM results with \textbf{BI} degradation. The best performance is shown in \textbf{bold}. The extension model achieves the best PSNR/SSIM results on all benchmarks.}
	\label{Tab:BI-result}
	\fontsize{6.5}{8}\selectfont
	\begin{tabular}{|c|c|c|c|c|c|c|c|c|c||c|}
		\hline  
		\textbf{Dataset}& Scale& Bicubic& SRCNN~\cite{srcnn_dong2015image}& VDSR~\cite{vdsr_kim2016accurate}& LapSRN~\cite{lapsrn}& EDSR~\cite{edsr_lim2017enhanced}& RDN~\cite{rdn_zhang2018residual}& SRFBN~\cite{srfbn_li2019feedback}& Ours& Ours+ \\
		\hline
		\hline
		\multirow{3}*{\textbf{Set5}}
		& $\times2$ & 33.66/0.9299 & 36.66/0.9542 & 37.53/0.9590 & 37.52/0.9591 & 38.11/0.9601 & 38.24/0.9614 & 38.11/0.9609 & 38.20/0.9613 & \textbf{38.25/0.9615}\\
		& $\times3$ & 30.39/0.8682 & 32.75/0.9090 & 33.67/0.9210 & 33.82/0.9227 & 34.65/0.9282 & 34.71/0.9296 & 34.70/0.9292 & 34.68/0.9294 & \textbf{34.76/0.9300}\\
		& $\times4$ & 28.42/0.8104 & 30.48/0.8628 & 31.35/0.8830 & 31.54/0.8850 & 32.46/0.8968 & 32.47/0.8990 & 32.47/0.8983 & 32.55/0.8992 & \textbf{32.66/0.9004}\\
		\hline
		\multirow{3}*{\textbf{Set14}}
		& $\times2$ & 30.24/0.8688 & 32.45/0.9067 & 33.05/0.9130 & 33.08/0.9130 & 33.92/0.9195 & 34.01/0.9212 & 33.82/0.9196 & 33.84/0.9199 & \textbf{34.03/0.9212}\\
		& $\times3$ & 27.55/0.7742 & 29.30/0.8215 & 29.78/0.8320 & 29.87/0.8320 & 30.52/0.8462 & 30.57/0.8468 & 30.51/0.8461 & 30.60/0.8475 & \textbf{30.67/0.8487}\\
		& $\times4$ & 26.00/0.7027 & 27.50/0.7513 & 28.02/0.7680 & 28.19/0.7720 & 28.80/0.7876 & 28.81/0.7871 & 28.81/0.7868 & 28.79/0.7872 & \textbf{28.91/0.7891}\\
		\hline
		\multirow{3}*{\textbf{B100}}
		& $\times2$ & 29.56/0.8431 & 31.36/0.8879 & 31.90/0.8960 & 31.80/0.8950 & 32.32/0.9013 & 32.34/0.9017 & 32.29/0.9010 & 32.35/0.9019 & \textbf{32.39/0.9023}\\
		& $\times3$ & 27.21/0.7385 & 28.41/0.7863 & 28.83/0.7990 & 28.82/0.7980 & 29.25/0.8093 & 29.26/0.8093 & 29.24/0.8084 & 29.25/0.8096 & \textbf{29.31/0.8105}\\
		& $\times4$ & 25.96/0.6675 & 26.90/0.7101 & 27.29/0.7260 & 27.32/0.7270 & 27.71/0.7420 & 27.72/0.7419 & 27.72/0.7409 & 27.74/0.7422 & \textbf{27.80/0.7435}\\
		\hline
		\multirow{3}*{\textbf{Urban100}}
		& $\times2$ & 26.88/0.8403 & 29.50/0.8946 & 30.77/0.9140 & 30.41/0.9101 & 32.93/0.9351 & 32.89/0.9353 & 32.62/0.9328 & 32.96/0.9357 & \textbf{33.10/0.9371}\\
		& $\times3$ & 24.46/0.7349 & 26.24/0.7989 & 27.14/0.8290 & 27.07/0.8280 & 28.80/0.8653 & 28.80/0.8653 & 28.73/0.8641 & 28.83/0.8666 & \textbf{29.01/0.8691}\\
		& $\times4$ & 23.14/0.6577 & 24.52/0.7221 & 25.18/0.7540 & 25.21/0.7560 & 26.64/0.8033 & 26.61/0.8028 & 26.60/0.8015 & 26.64/0.8033 & \textbf{26.83/0.8070}\\
		\hline
		\multirow{3}*{\textbf{Manga109}}
		& $\times2$ & 30.80/0.9339 & 35.60/0.9663 & 37.22/0.9750 & 37.27/0.9740 & 39.10/0.9773 & 39.18/0.9780 & 39.08/0.9779 & 39.20/0.9781 & \textbf{39.38/0.9785}\\
		& $\times3$ & 26.95/0.8556 & 30.48/0.9117 & 32.01/0.9340 & 32.21/0.9350 & 34.17/0.9476 & 34.13/0.9484 & 34.18/0.9481 & 34.19/0.9487 & \textbf{34.45/0.9499}\\
		& $\times4$ & 24.89/0.7866 & 27.58/0.8555 & 28.83/0.8870 & 29.09/0.8900 & 31.02/0.9148 & 31.00/0.9151 & 31.15/0.9160 & 31.16/0.9166 & \textbf{31.48/0.9190}\\
		\hline
		\hline
		\multicolumn{2}{|c|}{\textbf{Param~(M)}}& -&  0.057&			0.665&		0.813&			43&				20&			3.6&			3.4&			3.4\\
		\hline
	\end{tabular}
\end{table*}

\begin{figure}[t]
	\centering
	\includegraphics[width=\linewidth]{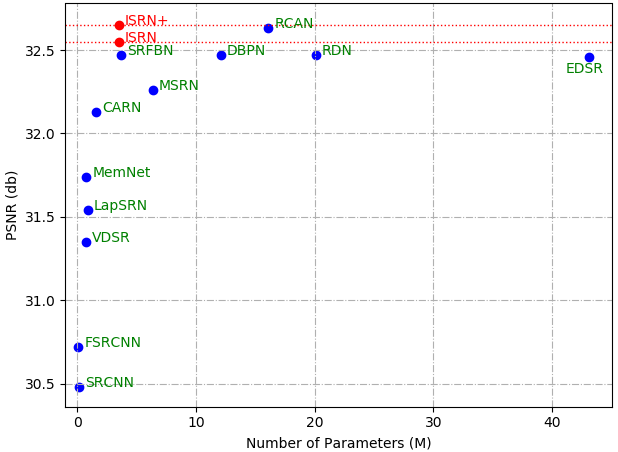}
	\caption{A visualization comparison of PSNR and parameters on Set5 with scaling factor $\times4$.}
	\label{Fig:set5-param}
\end{figure}

\subsection{Results with BI Degradation}
The experiments are conducted with \textbf{BI} (scaling factors $\times2$, $\times3$, and $\times4$). In particular, ISRN and ISRN$^+$ are compared with several methods, and Table~\ref{Tab:BI-result} shows quantitative comparisons. From the results, ISRN$^+$ achieves the best performance on all benchmark datasets, and ISRN achieves better performance than others on Urban100 and Manga109. Moreover, ISRN and ISRN$^+$ are superior in terms of SSIM values, which implies that the models can recover the structural information more effectively, as shown on B100, Urban100 and Manga109 datasets. Results on Urban100 and Manga109 show the performance on recovering structure information. A visualization comparison of PSNR and parameters on Set5~$\times4$ is shown in Fig.~\ref{Fig:set5-param}, which reveals that the proposed model achieves competitive results with fewer parameters than state-of-the-arts. 	

\begin{figure*}[t]
	\captionsetup[subfloat]{labelformat=empty, justification=centering}
	\begin{center}
		\newcommand{\rowArg}{1.5cm}
		\newcommand{\fullSize}{2.6cm}
		\newcommand{\patchSize}{2.0cm}
		\scriptsize
		\setlength\tabcolsep{0.1cm}
		\begin{tabular}[b]{c@{\hspace{0.1cm}}c@{\hspace{0.1cm}}c@{\hspace{0.1cm}}c@{\hspace{0.1cm}}c@{\hspace{0.1cm}}c@{\hspace{0.1cm}}c@{\hspace{0.1cm}}c}
			\subfloat[Image 8023]
			{\includegraphics[width = \fullSize, height = \patchSize]
				{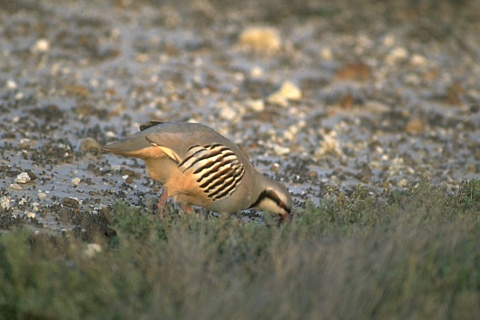}} &
			\subfloat[HR \protect\linebreak(PSNR/SSIM)]
			{\includegraphics[width = \patchSize, height = \patchSize]
				{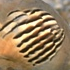}} &
			\subfloat[VDSR~\cite{vdsr_kim2016accurate}    \protect\linebreak(29.54/0.8651)]
			{\includegraphics[width = \patchSize, height = \patchSize]
				{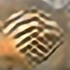}} &
			\subfloat[LapSRN~\cite{lapsrn}  \protect\linebreak(30.09/0.8670)]
			{\includegraphics[width = \patchSize, height = \patchSize]
				{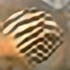}} &
			\subfloat[RDN~\cite{rdn_zhang2018residual}     \protect\linebreak(30.33/0.8760)]
			{\includegraphics[width = \patchSize, height = \patchSize]
				{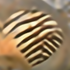}} &
			\subfloat[CARN~\cite{carn_ahn2018fast}   \protect\linebreak(30.94/0.8755)]
			{\includegraphics[width = \patchSize, height = \patchSize]
				{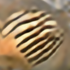}} &
			\subfloat[Ours    \protect\linebreak(31.08/0.8785)]
			{\includegraphics[width = \patchSize, height = \patchSize]
				{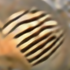}} &
			\subfloat[Ours+   \protect\linebreak(\textbf{31.18}/\textbf{0.8793})]
			{\includegraphics[width = \patchSize, height = \patchSize]
				{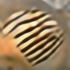}}
			
			\\[-0.3cm]
			
			\subfloat[Image 223061]
			{\includegraphics[width = \fullSize, height = \patchSize]
				{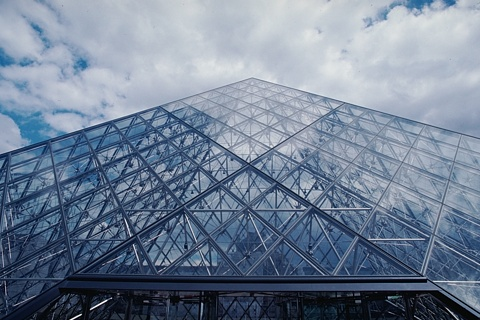}} &
			\subfloat[HR \protect\linebreak(PSNR/SSIM)]
			{\includegraphics[width = \patchSize, height = \patchSize]
				{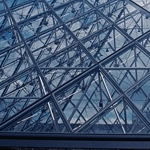}} &
			\subfloat[VDSR~\cite{vdsr_kim2016accurate}    \protect\linebreak(24.46/0.6537)]
			{\includegraphics[width = \patchSize, height = \patchSize]
				{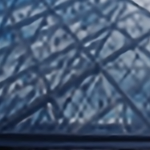}} &
			\subfloat[LapSRN~\cite{lapsrn}  \protect\linebreak(24.51/0.6566)]
			{\includegraphics[width = \patchSize, height = \patchSize]
				{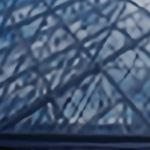}} &
			\subfloat[RDN~\cite{rdn_zhang2018residual}     \protect\linebreak(25.01/0.7032)]
			{\includegraphics[width = \patchSize, height = \patchSize]
				{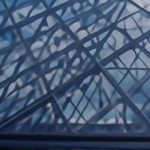}} &
			\subfloat[CARN~\cite{carn_ahn2018fast}   \protect\linebreak(24.75/0.6745)]
			{\includegraphics[width = \patchSize, height = \patchSize]
				{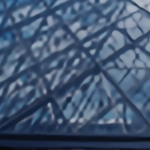}} &
			\subfloat[Ours    \protect\linebreak(25.07/0.7045)]
			{\includegraphics[width = \patchSize, height = \patchSize]
				{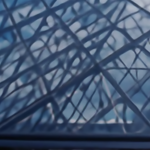}} &
			\subfloat[Ours+   \protect\linebreak(\textbf{25.12}/\textbf{0.7058})]
			{\includegraphics[width = \patchSize, height = \patchSize]
				{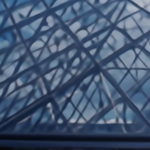}}
			
			\\[-0.3cm]
			
			\subfloat[Image 253027]
			{\includegraphics[width = \fullSize, height = \patchSize]
				{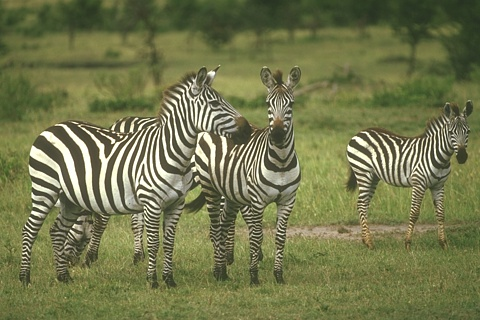}} &
			\subfloat[HR \protect\linebreak(PSNR/SSIM)]
			{\includegraphics[width = \patchSize, height = \patchSize]
				{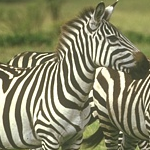}} &
			\subfloat[VDSR~\cite{vdsr_kim2016accurate}    \protect\linebreak(22.35/0.6902)]
			{\includegraphics[width = \patchSize, height = \patchSize]
				{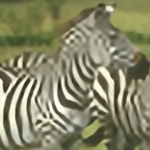}} &
			\subfloat[LapSRN~\cite{lapsrn}  \protect\linebreak(22.40/0.6922)]
			{\includegraphics[width = \patchSize, height = \patchSize]
				{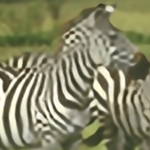}} &
			\subfloat[RDN~\cite{rdn_zhang2018residual}     \protect\linebreak(22.85/0.7118)]
			{\includegraphics[width = \patchSize, height = \patchSize]
				{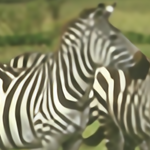}} &
			\subfloat[CARN~\cite{carn_ahn2018fast}   \protect\linebreak(22.67/0.7043)]
			{\includegraphics[width = \patchSize, height = \patchSize]
				{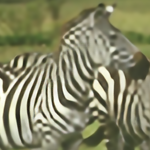}} &
			\subfloat[Ours    \protect\linebreak(22.93/0.7143)]
			{\includegraphics[width = \patchSize, height = \patchSize]
				{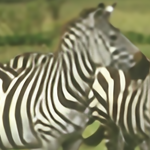}} &
			\subfloat[Ours+   \protect\linebreak(\textbf{23.12}/\textbf{0.7163})]
			{\includegraphics[width = \patchSize, height = \patchSize]
				{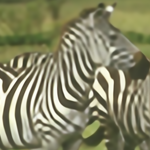}}
		\end{tabular}
	\end{center}
	\setlength{\abovecaptionskip}{0pt plus 2pt minus 2pt}
	\setlength{\belowcaptionskip}{0pt plus 2pt minus 2pt}
	\caption{Visual quality comparisons on B100 dataset with \textbf{BI}$\times4$ degradation.}
	\label{Fig:vis-result-bi-b100}
\end{figure*}

The visual quality comparisons on B100 dataset are shown in Fig.~\ref{Fig:vis-result-bi-b100}, which contains abundant complex structural textures from real world. From these results, ISRN and ISRN$^+$ can recover structural information more effectively. This also explains why the models can achieve promising SSIM result. When processing structural information, especially the line textures, ISRN and ISRN$^+$ have shown very competitive performance.

\begin{figure*}[t]
	\captionsetup[subfloat]{labelformat=empty, justification=centering}
	\begin{center}
		\newcommand{\rowArg}{1.5cm}
		\newcommand{\fullSize}{2.6cm}
		\newcommand{\patchSize}{2.0cm}
		\scriptsize
		\setlength\tabcolsep{0.1cm}
		\begin{tabular}[b]{c@{\hspace{0.1cm}}c@{\hspace{0.1cm}}c@{\hspace{0.1cm}}c@{\hspace{0.1cm}}c@{\hspace{0.1cm}}c@{\hspace{0.1cm}}c@{\hspace{0.1cm}}c}
			\subfloat[Image img\_062]
			{\includegraphics[width = \fullSize, height = \patchSize]
				{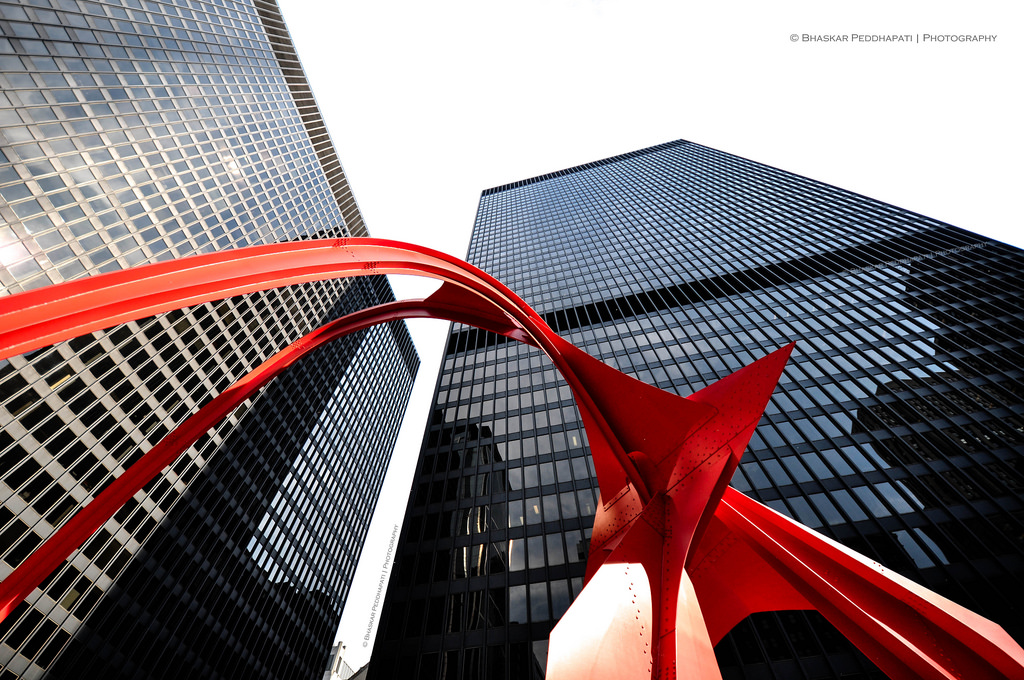}} &
			\subfloat[HR \protect\linebreak(PSNR/SSIM)]
			{\includegraphics[width = \patchSize, height = \patchSize]
				{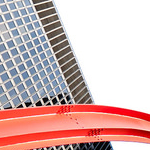}} &
			\subfloat[VDSR    \protect\linebreak(20.75/0.7474)]
			{\includegraphics[width = \patchSize, height = \patchSize]
				{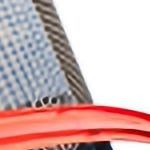}} &
			\subfloat[LapSRN  \protect\linebreak(20.80/0.7500)]
			{\includegraphics[width = \patchSize, height = \patchSize]
				{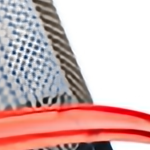}} &
			\subfloat[RDN     \protect\linebreak(22.31/0.8401)]
			{\includegraphics[width = \patchSize, height = \patchSize]
				{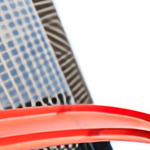}} &
			\subfloat[CARN   \protect\linebreak(21.39/0.7969)]
			{\includegraphics[width = \patchSize, height = \patchSize]
				{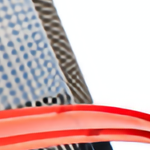}} &
			\subfloat[Ours    \protect\linebreak(22.41/0.8412)]
			{\includegraphics[width = \patchSize, height = \patchSize]
				{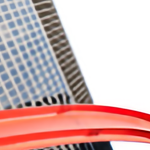}} &
			\subfloat[Ours+   \protect\linebreak(\textbf{22.53}/\textbf{0.8439})]
			{\includegraphics[width = \patchSize, height = \patchSize]
				{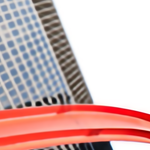}}
			
			\\[-0.3cm]
			
			\subfloat[Image img\_069]
			{\includegraphics[width = \fullSize, height = \patchSize]
				{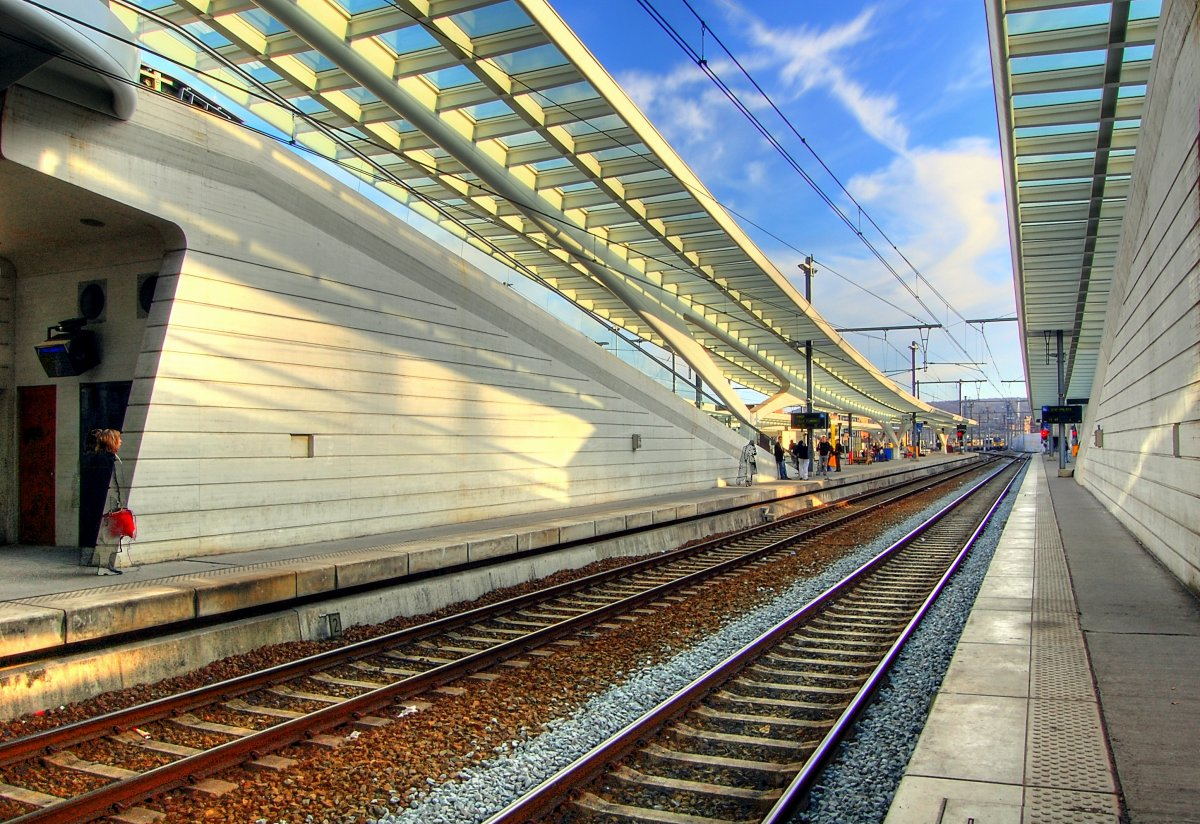}} &
			\subfloat[HR \protect\linebreak(PSNR/SSIM)]
			{\includegraphics[width = \patchSize, height = \patchSize]
				{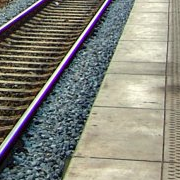}} &
			\subfloat[VDSR    \protect\linebreak(24.40/0.7320)]
			{\includegraphics[width = \patchSize, height = \patchSize]
				{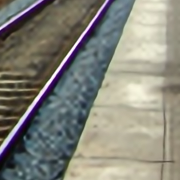}} &
			\subfloat[LapSRN  \protect\linebreak(24.39/0.7345)]
			{\includegraphics[width = \patchSize, height = \patchSize]
				{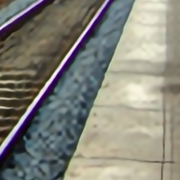}} &
			\subfloat[RDN     \protect\linebreak(25.19/0.7732)]
			{\includegraphics[width = \patchSize, height = \patchSize]
				{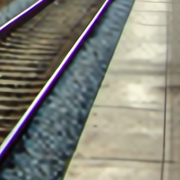}} &
			\subfloat[CARN   \protect\linebreak(24.78/0.7537)]
			{\includegraphics[width = \patchSize, height = \patchSize]
				{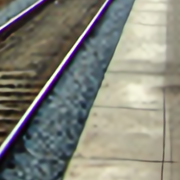}} &
			\subfloat[Ours    \protect\linebreak(25.23/0.7744)]
			{\includegraphics[width = \patchSize, height = \patchSize]
				{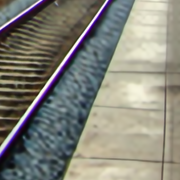}} &
			\subfloat[Ours+   \protect\linebreak(\textbf{25.28}/\textbf{0.7752})]
			{\includegraphics[width = \patchSize, height = \patchSize]
				{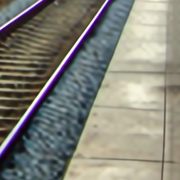}}
			
			\\[-0.3cm]
			
						\subfloat[Image img\_070]
						{\includegraphics[width = \fullSize, height = \patchSize]
							{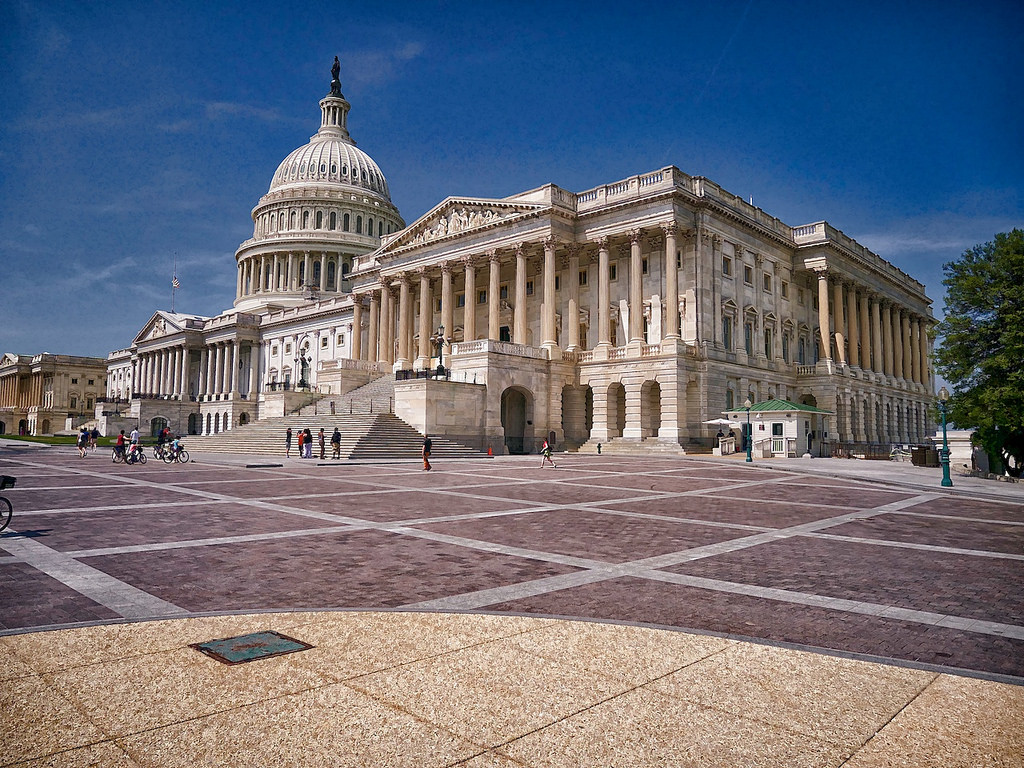}} &
						\subfloat[HR \protect\linebreak(PSNR/SSIM)]
						{\includegraphics[width = \patchSize, height = \patchSize]
							{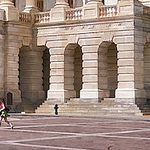}} &
						\subfloat[VDSR    \protect\linebreak(21.92/0.5767)]
						{\includegraphics[width = \patchSize, height = \patchSize]
							{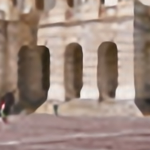}} &
						\subfloat[LapSRN  \protect\linebreak(21.93/0.5776)]
						{\includegraphics[width = \patchSize, height = \patchSize]
							{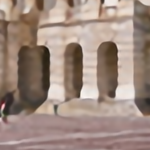}} &
						\subfloat[RDN     \protect\linebreak(22.20/0.6070)]
						{\includegraphics[width = \patchSize, height = \patchSize]
							{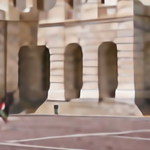}} &
						\subfloat[CARN   \protect\linebreak(22.12/0.5936)]
						{\includegraphics[width = \patchSize, height = \patchSize]
							{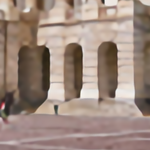}} &
						\subfloat[Ours    \protect\linebreak(22.35/0.6100)]
						{\includegraphics[width = \patchSize, height = \patchSize]
							{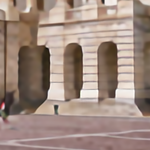}} &
						\subfloat[Ours+   \protect\linebreak(\textbf{22.37}/\textbf{0.6110})]
						{\includegraphics[width = \patchSize, height = \patchSize]
							{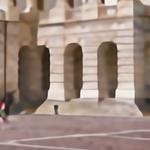}}
						
						\\[-0.3cm]
			
			\subfloat[Image img\_096]
			{\includegraphics[width = \fullSize, height = \patchSize]
				{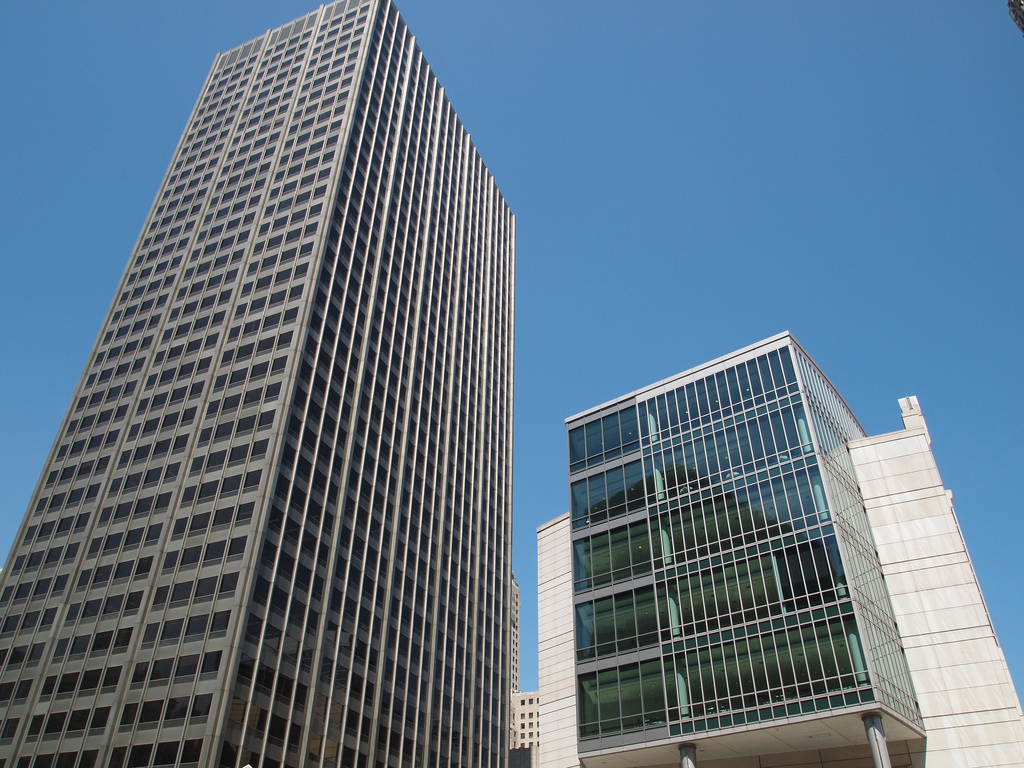}} &
			\subfloat[HR \protect\linebreak(PSNR/SSIM)]
			{\includegraphics[width = \patchSize, height = \patchSize]
				{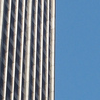}} &
			\subfloat[VDSR    \protect\linebreak(23.31/0.8014)]
			{\includegraphics[width = \patchSize, height = \patchSize]
				{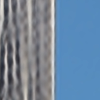}} &
			\subfloat[LapSRN  \protect\linebreak(22.53/0.7851)]
			{\includegraphics[width = \patchSize, height = \patchSize]
				{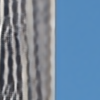}} &
			\subfloat[RDN     \protect\linebreak(26.14/0.8921)]
			{\includegraphics[width = \patchSize, height = \patchSize]
				{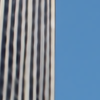}} &
			\subfloat[CARN   \protect\linebreak(25.11/0.8629)]
			{\includegraphics[width = \patchSize, height = \patchSize]
				{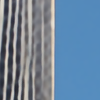}} &
			\subfloat[Ours    \protect\linebreak(26.63/0.9849)]
			{\includegraphics[width = \patchSize, height = \patchSize]
				{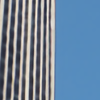}} &
			\subfloat[Ours+   \protect\linebreak(\textbf{27.10}/\textbf{0.9006})]
			{\includegraphics[width = \patchSize, height = \patchSize]
				{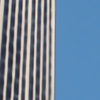}}
		\end{tabular}
	\end{center}
	\setlength{\abovecaptionskip}{0pt plus 2pt minus 2pt}
	\setlength{\belowcaptionskip}{0pt plus 2pt minus 2pt}
	\caption{Visual quality comparisons on Urban100 dataset with \textbf{BI}$\times4$ degradation.}
	\label{Fig:vis-result-bi-urban100}
\end{figure*}

To show the performance on large images with more textures, we compare the models with other works on Urban100 dataset. The images are from urban photos, which contain more line and structural textures. Visualization quality comparisons on Urban100 dataset are shown in Fig.~\ref{Fig:vis-result-bi-urban100}. Compared with RDN, ISRN and ISRN$^+$ could recover more textures on buildings. Specifically, our models can distinguish the mixture of lines more efficiently.

We also compare the computation complexity with state-of-the-arts. For a fair comparison, the computation complexity is modeled as the number of multiply-accumulate operations~(MACs). As a hardware-free indicator, MACs can well describe the mathematical complexity of every model. MACs is calculated by restoring a 720P $(1280\times720)$ image with $\textbf{BI}\times4$ degradation. Table~\ref{Tab:flops} shows the parameter, MACs, and performance comparisons. ISRN achieves competitive or better restoration performance with much fewer MACs and parameters, which shows the effectiveness of the proposed iterative design.

\begin{table}[t]
    \centering
    \caption{Perceptual and quality comparison with GAN-based method with $\textbf{BI}\times4$ degradation.}
    \label{Tab:esrgan}
    \begin{tabular}{|c|c|c|c|c|}
         \hline
         \textbf{Dataset}& \multicolumn{2}{c|}{\textbf{Set5}}& \multicolumn{2}{c|}{\textbf{Set14}} \\
         \hline
         \textbf{Model}& ISRN& ESRGAN& ISRN& ESRGAN \\
         \hline
         \textbf{PSNR}& 32.55& 30.45& 28.79& 26.28 \\
         \textbf{SSIM}& 0.8992& 0.8516& 0.7872& 0.6980 \\
         \textbf{RMSE}& 6.4200& 8.0834& 11.0051& 15.0205 \\
         \textbf{PI}& 5.8996& 3.7768& 5.2871& 2.9188 \\
         \hline
    \end{tabular}
\end{table}

Furthermore, we compare ISRN with GAN-based methods. By adopting an end-to-end training strategy, all components in ISRN are trained directly with $\ell_1$-loss between $\mathbf{I}^{SR}$ and $\mathbf{I}^{HR}$, and no extra or specific criterion is considered. Different from GAN-based methods that encourage high frequency crispness effectively, the pixel-wise loss ensures the reliable textures with no additional generated contents. To demonstrate the restoration capacity, we compare ISRN with ESRGAN~\cite{esrgan} by using both subjective and objective indicators~\cite{pirm}. Table~\ref{Tab:esrgan} shows that ISRN can achieve better objective results than GAN-based method. Specially, ISRN can achieve more than 2 dB improvement on PSNR. Since there is no GAN-based loss, the perceptual performances of ISRN are not better than ESRGAN. The higher PSNR/SSIM results demonstrate the results of ISRN are credible.

\begin{table*}[t]
	\centering
	\caption{Average PSNR/SSIM results with \textbf{BD} and \textbf{DN} degradation. The best performance is shown in \textbf{bold}. The basic and extension models achieve better PSNR/SSIM results on all benchmarks than state-of-the-arts.}
	\label{Tab:BD-result}
	\fontsize{6.5}{8}\selectfont
	\begin{tabular}{|c|c|c|c|c|c|c|c|c|c||c|}
		\hline
		\textbf{Dataset}& Scale& Bicubic& SRCNN~\cite{srcnn_dong2015image}& IRCNN\_G~\cite{ircnn_zhang2017learning}& IRCNN\_C~\cite{ircnn_zhang2017learning}& RDN~\cite{rdn_zhang2018residual}& RCAN~\cite{rcan_zhang2018image}& SRFBN~\cite{srfbn_li2019feedback}& Ours& Ours+ \\
		\hline
		\hline
		\multirow{2}*{\textbf{Set5}}
		& \textbf{BD} & 28.34/0.8161 & 31.63/0.8888 & 33.38/0.9182 & 29.55/0.8246 & 34.57/0.9280 & 34.70/0.9288 & 34.66/0.9283 & 34.74/0.9291 & \textbf{34.83/0.9297} \\
		& \textbf{DN} & 24.14/0.5445 & 27.16/0.7672 & 24.85/0.7205 & 26.18/0.7430 & 28.46/0.8151 &-& 28.53/0.8182 & 28.59/0.8201 & \textbf{28.66/0.8214} \\
		\hline
		\multirow{2}*{\textbf{Set14}}
		& \textbf{BD} & 26.12/0.7106 & 28.52/0.7924 & 29.73/0.8292 & 27.33/0.7135 & 30.53/0.8447 & 30.63/0.8462 & 30.48/0.8439 & 30.69/0.8473 & \textbf{30.78/0.8484} \\
		& \textbf{DN} & 23.14/0.4828 & 25.49/0.6580 & 23.84/0.6091 & 24.68/0.6300 & 26.60/0.7101 &-& 26.60/0.7144 & 26.71/0.7167 & \textbf{26.75/0.7175} \\
		\hline
		\multirow{2}*{\textbf{B100}}
		& \textbf{BD} & 26.02/0.6733 & 27.76/0.7526 & 28.65/0.7922 & 26.46/0.6572 & 29.23/0.8079 & 29.32/0.8093 & 29.21/0.8069 & 29.31/0.8099 & \textbf{29.36/0.8107} \\
		& \textbf{DN} & 22.94/0.4461 & 25.11/0.6151 & 23.89/0.5688 & 24.52/0.5850 & 25.93/0.6573 &-& 25.95/0.6625 & 26.00/0.6637 & \textbf{26.03/0.6644} \\
		\hline
		\multirow{2}*{\textbf{Urban100}}
		& \textbf{BD} & 23.20/0.6661 & 25.31/0.7612 & 26.77/0.8154 & 24.89/0.7172 & 28.46/0.8581 & 28.81/0.8647 & 28.48/0.8581 & 28.83/0.8652 & \textbf{29.01/0.8680} \\
		& \textbf{DN} & 21.63/0.4701 & 23.32/0.6500 & 21.96/0.6018 & 22.63/0.6205 & 24.92/0.7362 &-& 24.99/0.7424 & 25.25/0.7525 & \textbf{25.35/0.7549} \\
		\hline
		\multirow{2}*{\textbf{Manga109}}
		& \textbf{BD} & 25.03/0.7987 & 28.79/0.8851 & 31.15/0.9245 & 28.68/0.8574 & 33.97/0.9465 & 34.38/0.9483 & 34.07/0.9466 & 34.46/0.9489 & \textbf{34.73/0.9501} \\
		& \textbf{DN} & 23.08/0.5448 & 25.78/0.7889 & 23.18/0.7466 & 24.74/0.7701 & 28.00/0.8590 &-& 28.02/0.8618 & 28.25/0.8669 & \textbf{28.39/0.8688} \\
		\hline
	\end{tabular}
\end{table*}

\begin{figure*}[t]
	\newcommand{\rowArg}{2.0cm}
	\captionsetup[subfloat]{labelformat=empty, justification=centering}
	\begin{center}
		\begin{tabular}[b]{c@{\hspace{0.1cm}}c@{\hspace{0.1cm}}c@{\hspace{0.1cm}}c@{\hspace{0.1cm}}c@{\hspace{0.1cm}}c@{\hspace{0.1cm}}c}
			\subfloat[img\_012 from Urban100]
			{\includegraphics[height=\rowArg]
				{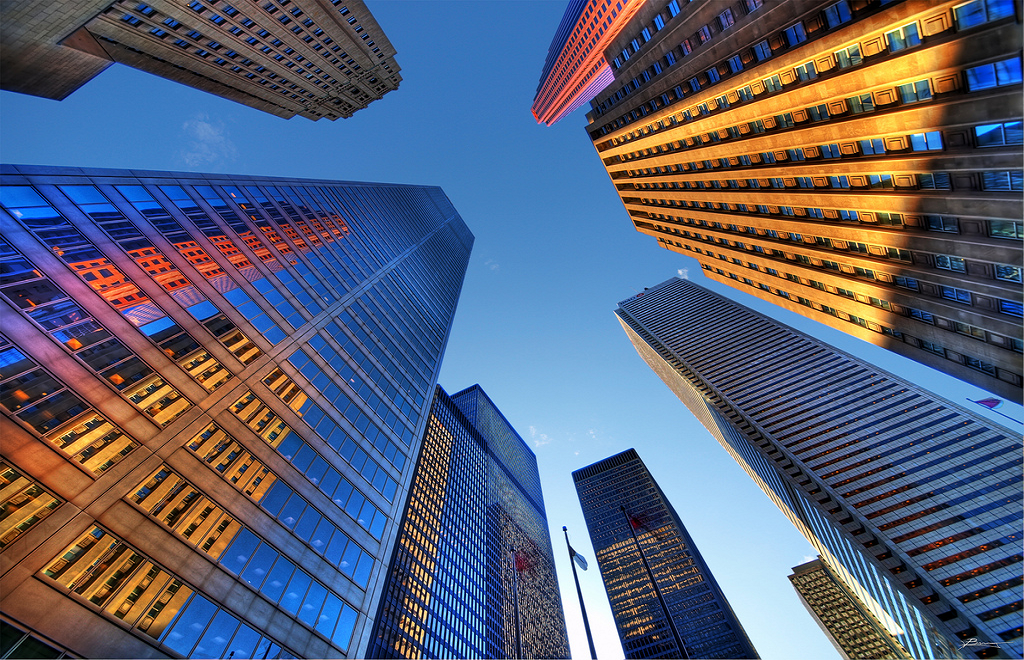}} &
			\subfloat[HR~\protect\linebreak(PSNR/SSIM)]
			{\includegraphics[width=\rowArg]
				{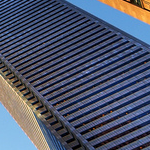}} &
			\subfloat[Bicubic~\protect\linebreak(22.52/0.6147)]
			{\includegraphics[width=\rowArg]
				{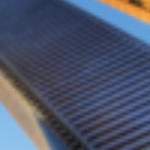}} &
			\subfloat[RDN~\protect\linebreak(25.53/0.8236)]
			{\includegraphics[width=\rowArg]
				{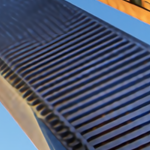}} &
			\subfloat[SRFBN~\protect\linebreak(25.44/0.8221)]
			{\includegraphics[width=\rowArg]
				{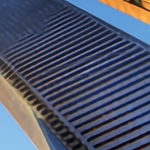}} &
			\subfloat[Ours~\protect\linebreak(26.18/0.8390)]
			{\includegraphics[width=\rowArg]
				{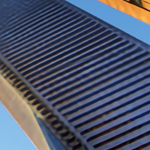}} &
			\subfloat[Ours+~\protect\linebreak(\textbf{26.29/0.8403})]
			{\includegraphics[width=\rowArg]
				{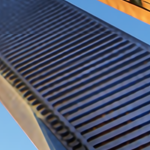}} \\[-0.3cm]
			\subfloat[img\_024 from Urban100]
			{\includegraphics[height=\rowArg]
				{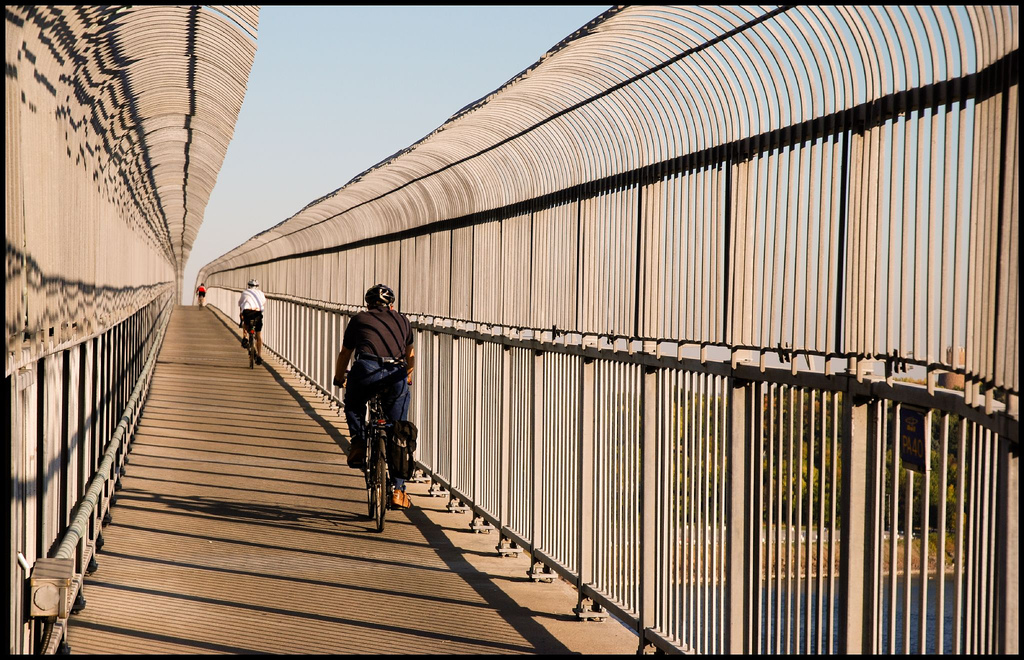}} &
			\subfloat[HR~\protect\linebreak(PSNR/SSIM)]
			{\includegraphics[width=\rowArg]
				{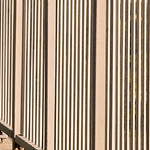}} &
			\subfloat[Bicubic~\protect\linebreak(18.41/0.4882)]
			{\includegraphics[width=\rowArg]
				{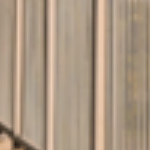}} &
			\subfloat[RDN~\protect\linebreak(22.85/0.7799)]
			{\includegraphics[width=\rowArg]
				{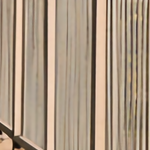}} &
			\subfloat[SRFBN~\protect\linebreak(22.92/0.7889)]
			{\includegraphics[width=\rowArg]
				{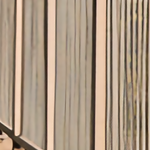}} &
			\subfloat[Ours~\protect\linebreak(23.58/0.8058)]
			{\includegraphics[width=\rowArg]
				{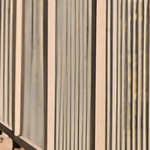}} &
			\subfloat[Ours+~\protect\linebreak(\textbf{23.89/0.8123})]
			{\includegraphics[width=\rowArg]
				{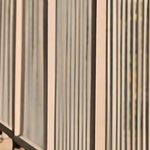}} \\[-0.3cm]
			\subfloat[img\_078 from Urban100]
			{\includegraphics[height=\rowArg]
				{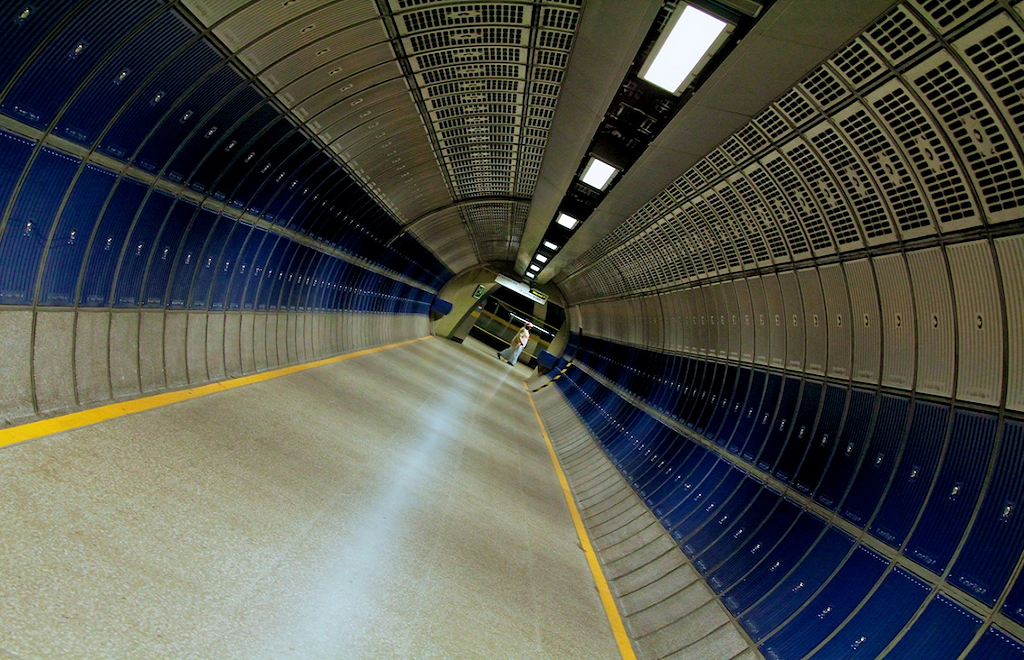}} &
			\subfloat[HR~\protect\linebreak(PSNR/SSIM)]
			{\includegraphics[width=\rowArg]
				{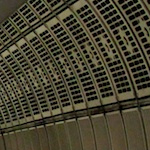}} &
			\subfloat[Bicubic~\protect\linebreak(25.73/0.6848)]
			{\includegraphics[width=\rowArg]
				{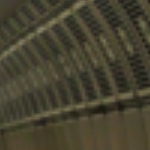}} &
			\subfloat[RDN~\protect\linebreak(29.92/0.8503)]
			{\includegraphics[width=\rowArg]
				{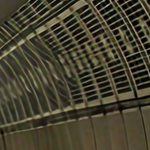}} &
			\subfloat[SRFBN~\protect\linebreak(29.82/0.8488)]
			{\includegraphics[width=\rowArg]
				{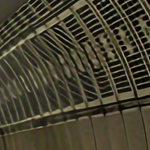}} &
			\subfloat[Ours~\protect\linebreak(30.62/0.8619)]
			{\includegraphics[width=\rowArg]
				{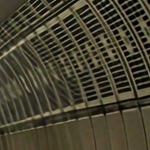}} &
			\subfloat[Ours+~\protect\linebreak(\textbf{30.75/0.8621})]
			{\includegraphics[width=\rowArg]
				{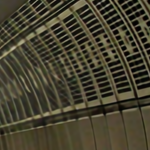}}
			\\[-0.3cm]
			\subfloat[86000 from B100]
			{\includegraphics[height=\rowArg]
				{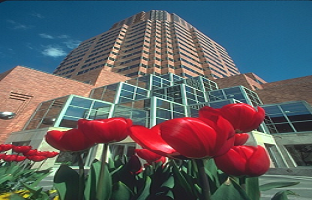}} &
			\subfloat[HR~\protect\linebreak(PSNR/SSIM)]
			{\includegraphics[width=\rowArg]
				{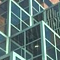}} &
			\subfloat[Bicubic~\protect\linebreak(24.17/0.6993)]
			{\includegraphics[width=\rowArg]
				{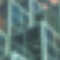}} &
			\subfloat[RDN~\protect\linebreak(28.70/0.8802)]
			{\includegraphics[width=\rowArg]
				{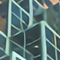}} &
			\subfloat[SRFBN~\protect\linebreak(28.72/0.8797)]
			{\includegraphics[width=\rowArg]
				{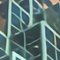}} &
			\subfloat[Ours~\protect\linebreak(29.39/0.8916)]
			{\includegraphics[width=\rowArg]
				{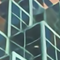}} &
			\subfloat[Ours+~\protect\linebreak(\textbf{29.53/0.8939})]
			{\includegraphics[width=\rowArg]
				{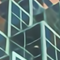}}
		\end{tabular}
	\end{center}
	\caption{Visual quality comparisons of different methods with \textbf{BD} degradation.}
	\label{Fig:vis-result-bd}
\end{figure*}

\begin{figure*}[t]
	\captionsetup[subfloat]{labelformat=empty, justification=centering}
	\newcommand{\rowArg}{2.0cm}
	\begin{center}
		\begin{tabular}[b]{c@{\hspace{0.1cm}}c@{\hspace{0.1cm}}c@{\hspace{0.1cm}}c@{\hspace{0.1cm}}c@{\hspace{0.1cm}}c@{\hspace{0.1cm}}c}
			\subfloat[img\_044 from Urban100]
			{\includegraphics[height=\rowArg]
				{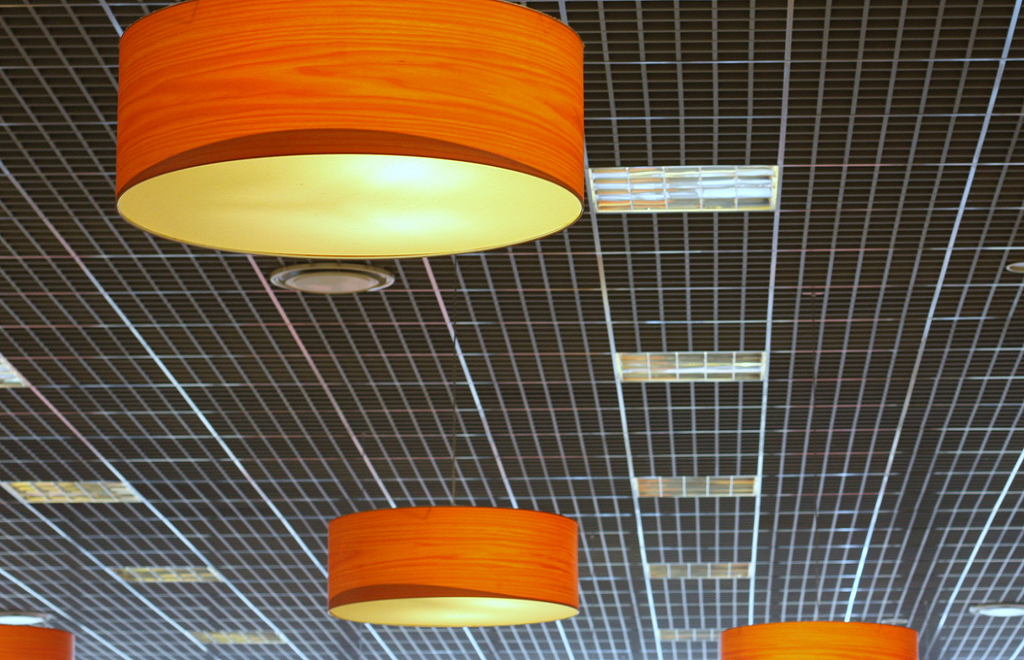}} &
			\subfloat[HR    \protect\linebreak(PSNR/SSIM)]
			{\includegraphics[width=\rowArg]
				{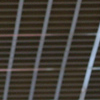}} &
			\subfloat[Bicubic   \protect\linebreak(24.58/0.5733)]
			{\includegraphics[width=\rowArg]
				{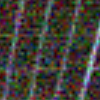}} &
			\subfloat[RDN     \protect\linebreak(28.35/0.7962)]
			{\includegraphics[width=\rowArg]
				{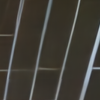}} &
			\subfloat[SRFBN   \protect\linebreak(29.26/0.8251)]
			{\includegraphics[width=\rowArg]
				{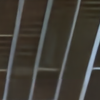}} &
			\subfloat[Ours    \protect\linebreak(\textcolor{red}{30.15}/\textcolor{red}{0.8604})]
			{\includegraphics[width=\rowArg]
				{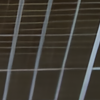}} &
			\subfloat[Ours+   \protect\linebreak(\textbf{30.43}/\textbf{0.8639})]
			{\includegraphics[width=\rowArg]
				{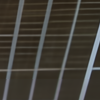}} \\[-0.3cm]
			\subfloat[img\_004 from Urban100]
			{\includegraphics[height=\rowArg]
				{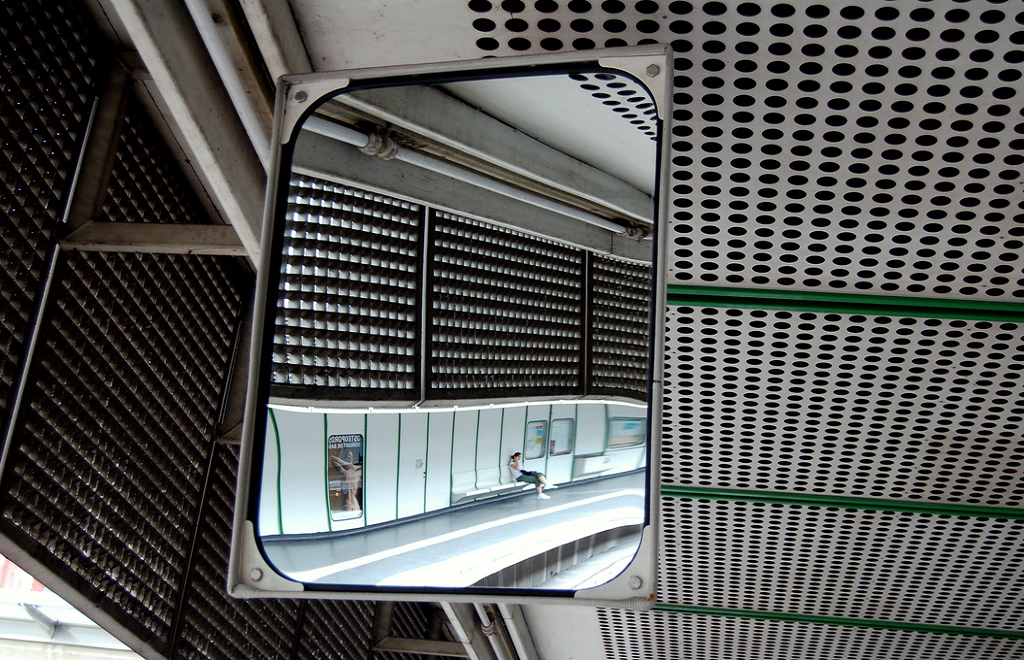}} &
			\subfloat[HR~\protect\linebreak(PSNR/SSIM)]
			{\includegraphics[width=\rowArg]
				{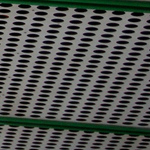}} &
			\subfloat[Bicubic~\protect\linebreak(20.78/0.5852)]
			{\includegraphics[width=\rowArg]
				{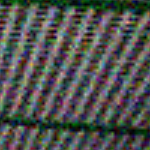}} &
			\subfloat[RDN~\protect\linebreak(23.27/0.8029)]
			{\includegraphics[width=\rowArg]
				{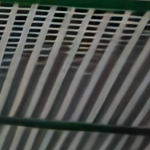}} &
			\subfloat[SRFBN~\protect\linebreak(23.00/0.7995)]
			{\includegraphics[width=\rowArg]
				{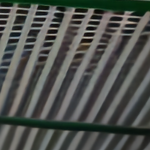}} &
			\subfloat[Ours~\protect\linebreak(23.76/0.8216)]
			{\includegraphics[width=\rowArg]
				{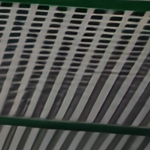}} &
			\subfloat[Ours+~\protect\linebreak(\textbf{24.08/0.8287})]
			{\includegraphics[width=\rowArg]
				{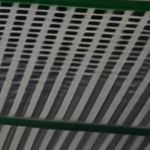}} \\[-0.3cm]
			\subfloat[img\_011 from Urban100]
			{\includegraphics[height=\rowArg]
				{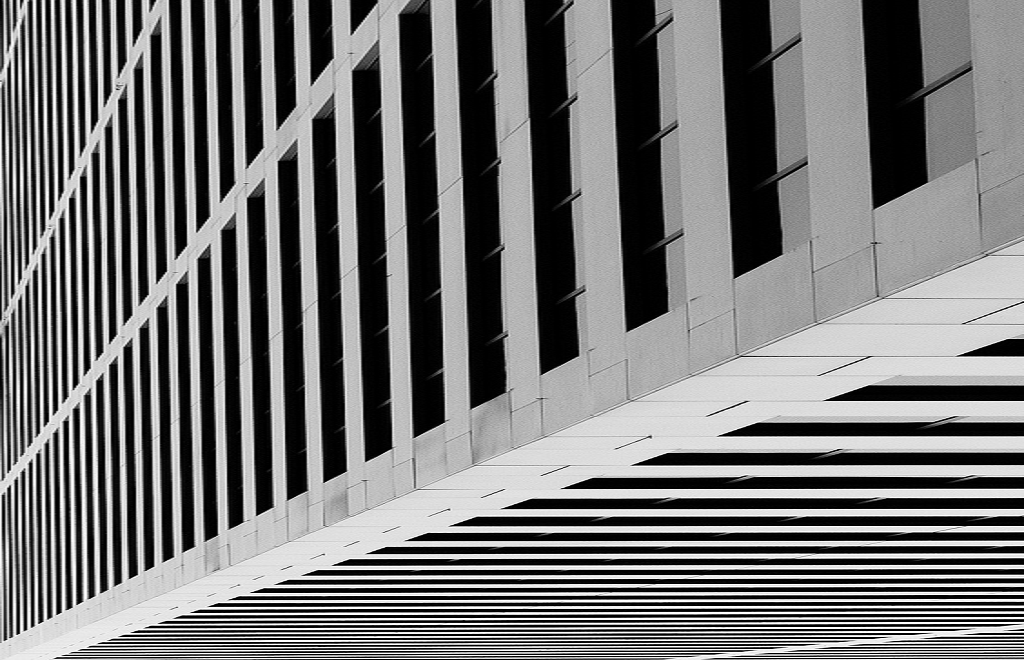}} &
			\subfloat[HR~\protect\linebreak(PSNR/SSIM)]
			{\includegraphics[width=\rowArg]
				{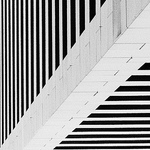}} &
			\subfloat[Bicubic~\protect\linebreak(16.66/0.5377)]
			{\includegraphics[width=\rowArg]
				{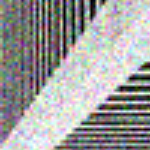}} &
			\subfloat[RDN~\protect\linebreak(19.08/0.8234)]
			{\includegraphics[width=\rowArg]
				{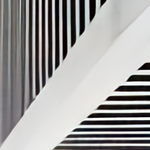}} &
			\subfloat[SRFBN~\protect\linebreak(19.90/0.8413)]
			{\includegraphics[width=\rowArg]
				{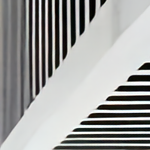}} &
			\subfloat[Ours~\protect\linebreak(20.68/0.8653)]
			{\includegraphics[width=\rowArg]
				{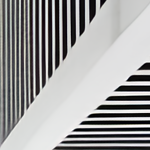}} &
			\subfloat[Ours+~\protect\linebreak(\textbf{20.91/0.8642})]
			{\includegraphics[width=\rowArg]
				{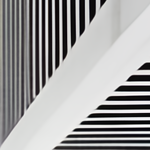}} \\ [-0.3cm]
		\end{tabular}
	\end{center}
	\caption{Visual quality comparisons of different methods with \textbf{DN} degradation.}
	\label{Fig:vis-result-dn}
\end{figure*}

\subsection{Results with BD and DN Degradation}
There are also experiments conducted with \textbf{BD} and \textbf{DN} degradation with the scaling factor $\times3$ to simulate the complex situations. Quantitative results are shown in Table~\ref{Tab:BD-result}. From the results, ISRN and ISRN$^+$ both achieve better performance than others. In particular, for the \textbf{DN} degradation, the proposed ISRN/ISRN$^+$ are superior in terms of both PSNR and SSIM. The promising performance is originated from the iterative structure which is the distinctive component compared with the prominent methods. 

The visual quality comparisons are shown in Fig.~\ref{Fig:vis-result-bd} and Fig.~\ref{Fig:vis-result-dn}. From Fig.~\ref{Fig:vis-result-bd} with \textbf{BD} degradation, the recovered lines from other works are warped or blurry. In ISRN and ISRN$^+$, the lines could be recovered better than other methods. From Fig.~\ref{Fig:vis-result-dn} with \textbf{DN} degradation, the tiny lines are missing from others work, due to the introduced noise. Random noise may disturb the original texture and make the tiny lines omitted. In ISRN and ISRN$^+$, the lines could be recovered better than other methods.

\subsection{Ablation Study}
\textbf{Study on Network Designs}. In the proposed ISRN, different \textit{Down-sampler} and \textit{Solver LR} are applied in different iterations, leading to better performance in general. The comparisons are performed with using same \textit{Down-sampler} and \textit{Solver LR} in different iterations. The results are shown in Table~\ref{Tab:ablation-diff}. From Table~\ref{Tab:ablation-diff}, the performance is better when using different components. Experimental results provide useful evidence regarding the training of different solvers. Since \textit{Solver LR} and \textit{Down-sampler} are simple components with restricted convolutional layers, they will lead to few increase of parameters.

\begin{table}[t]
	\centering
	\caption{PSNR/SSIM results with same and different solvers with \textbf{BI}~$\times4$ degradation.}
	\label{Tab:ablation-diff}
	\fontsize{9}{9}\selectfont
	\begin{tabular}{|c|c|c|c|}
		\hline
		\textbf{Solvers}	 	&Set5	 		&Set14			&B100\\
		\hline
		\hline
		\textbf{Same}		&32.43/0.8980	&28.80/0.7870	&27.70/0.7409\\
		\textbf{Different}	&32.55/0.8992	&28.79/0.7872	&27.74/0.7422\\
		\hline
	\end{tabular}
\end{table}

\begin{figure}[t]
	\centering
	\includegraphics[width=0.8\linewidth]{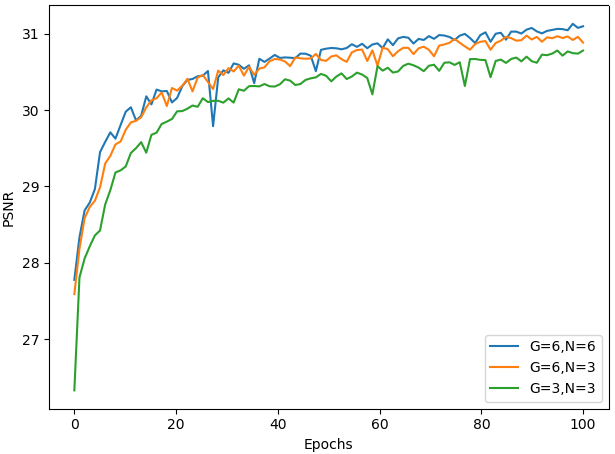}
	\caption{PSNR comparisons on different blocks and groups with \textbf{BI}~$\times4$ degradation. }
	\label{Fig:ablation-silm}
\end{figure}

The number of blocks will also influence the performance. There are experiments with different block number $N$ and group number $G$ to show the performance with different number of blocks and groups. The results are shown in Fig.~\ref{Fig:ablation-silm}. The comparisons are conducted on five validation images from DIV2K. From Fig.~\ref{Fig:ablation-silm}, the performance will be better with the increase of $N$ and $G$, showing more blocks can achieve better performance.

To demonstrate the effectiveness of padding structure, we compare the performance with $\textbf{BI}\times4$ degradation. Table~\ref{Tab:padding} shows the PSNR/SSIM results. From the comparison, one can be observed that padding boosts the restoration performance. Since it is a simple structure with restricted layers, applying padding structure leads to less complexity increase and better performance.

\begin{table}[t]
    \centering
    \caption{Performance comparison on padding structure with $\textbf{BI}\times4$ degradation.}
    \label{Tab:padding}
    \begin{tabular}{|c|c|c|c|}
        \hline
         \textbf{Padding}& B100& Urban100& Manga109\\
         \hline
         \hline
         \textbf{w}& 27.74/0.7422& 26.64/0.8033& 31.16/0.9166 \\
         \textbf{w/o}& 27.73/0.7421& 26.62/0.8030& 31.12/0.9164 \\
         \hline
    \end{tabular}
\end{table}

\textit{Solver MLE} is another important component for solving the memory-less issue. Since \textit{Solver HR} and \textit{Down-Sampler} adaptively learn the mapping between the LR and HR spaces, the memory-less situation (without MLE) is similar to ISRN with $K=1$ and only one available direction is considered. To address this point, Table~\ref{Tab:mle} shows the performances from two situations, where the methods are updated for 200 epochs. From the comparison, the two situations hold competitive performances for restoration.

\begin{table}[t]
    \centering
    \caption{Performance comparison on \textit{Solver MLE} with $\textbf{BI}\times4$ degradation.}
    \label{Tab:mle}
    \begin{tabular}{|c|c|c|c|}
         \hline
         \textbf{Method}& B100& Urban100& Manga109 \\
         \hline
         \hline
         \textbf{w/o MLE}& 27.55/0.7352& 26.00/0.7839& 30.31/0.9050 \\
         \textbf{ISRN}($K=1$)& 27.57/0.7364& 26.08/0.7864& 30.38/0.9072 \\
         \hline
    \end{tabular}
\end{table}

\textbf{Study on Feature Normalization}. To show the performance of feature normalization, experiments are conducted on Set5. The results are shown in Table~\ref{Tab:ablation-adafm}. From the results, the model with feature normalization achieves better performance with \textbf{BI} degradation (scaling factors $\times2$, $\times3$, and $\times4$). Since feature normalization is an elaborate and effective component, introducing the block will lead to few increases of parameters and computational cost. 

\begin{table}[t]
	\centering
	\caption{PSNR/SSIM results with and without feature normalization on Set5 with \textbf{BI} degradation.}
	\label{Tab:ablation-adafm}
	\fontsize{9}{9}\selectfont
	\begin{tabular}{|c|c|c|c|}
		\hline
		\textbf{FN} 	&	$\times2$ 		&	$\times3$		&	$\times4$ 		\\
		\hline
		\hline
		\textbf{w} 			&	38.20/0.9613	&	34.68/0.9294	&	32.55/0.8992	\\
		\textbf{w/o} 		&	38.18/0.9611	&	34.59/0.9289	&	32.48/0.8986	\\
		\hline
	\end{tabular}
\end{table}

Furthermore, we compare FN with BN. Table~\ref{Tab:fnbn} shows the performance comparison with $\textbf{BI}\times4$ degradation. From the results, FN achieves better restoration performance than BN.

\begin{table}[t]
    \centering
    \caption{Performance comparison on BN with \textbf{BI}$\times4$ degradation.}
    \label{Tab:fnbn}
    \begin{tabular}{|c|c|c|c|}
         \hline
         \textbf{Norm}& Set5& Set14& B100 \\
         \hline \hline
         \textbf{FN}& 32.55/0.8992& 28.79/0.7872& 27.74/0.7422 \\
         \textbf{BN}& 30.74/0.8697& 27.65/0.7589& 26.95/0.7143 \\
         \hline
    \end{tabular}
\end{table}

\begin{figure}[t]
	\centering
	\includegraphics[width=0.8\linewidth]{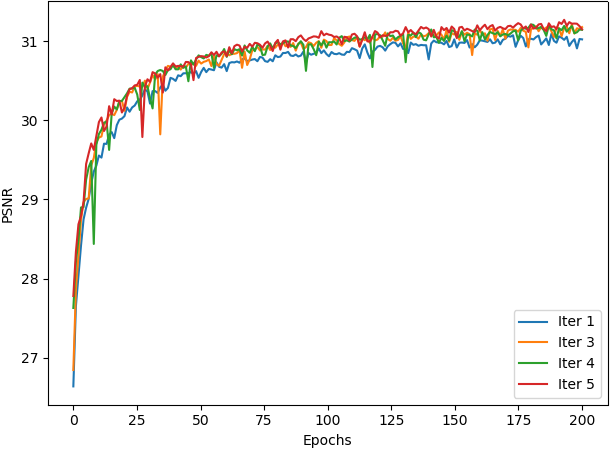}
	\caption{Performance comparisons for different iteration times with \textbf{BI}~$\times4$ degradation.}
	\label{Fig:ablation-iter}
\end{figure}

\textbf{Study on Iteration Mechanism}. To show the performance of iterations, experiment results with and without iterations are compared. The comparison is set with iteration number $k=1, 3, 4, 5$. The experiments on 5 validation images from DIV2K with scaling factor $\times4$ are conducted. The results are shown in Fig.~\ref{Fig:ablation-iter}. From the results, it could be found that iterations indeed improve the performance. With the increase of iteration times, the PSNR results will be higher.

For furthermore exploration on iteration mechanism, we compare the ISRN without iteration ($K=1$) with other works. Table~\ref{Tab:flops} demonstrates the parameters, MACs, and PSNR/SSIM results. From the results, ISRN ($K=1$) achieves competitive performance than other works with fewer parameters and less computation complexity. Based on the effective design, ISRN achieves a large performance improvement with the increase of $K$.

With the increase of $K$, the restoration performance will be improved. However, a larger $K$ will lead to higher computational complexity. Fig.~\ref{Fig:ablation-iter} shows that the improvement is restricted from $K=4$ to $5$. From this point of view, $K=5$ may be a suitable setting for balancing the performance and speed. Since DIV2K~\cite{div2k_timofte2017ntire} is a high-resolution dataset with diverse textures, the setting of $K$ may be suitable for general benchmarks.

\begin{figure}
	\vspace{-0.2cm}
	\captionsetup[subfloat]{labelformat=empty}
	\begin{center}
		\newcommand{\patchSize}{0.12\textwidth}
		\scriptsize
		\setlength\tabcolsep{0.1cm}
		\begin{tabular}[b]{ccc}
			\subfloat[(a)]
			{\includegraphics[width = \patchSize, height = \patchSize]
				{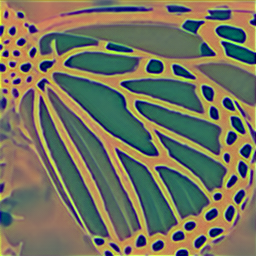}} &
			\subfloat[(b)]
			{\includegraphics[width = \patchSize, height = \patchSize]
				{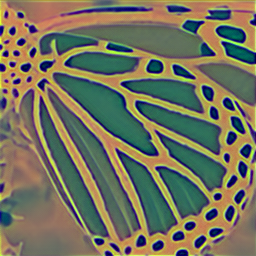}} &
			\subfloat[(c)]
			{\includegraphics[width = \patchSize, height = \patchSize]
				{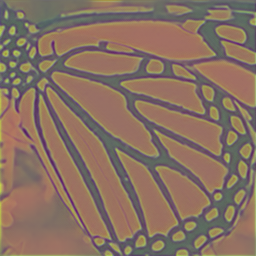}} \\ [-0.35cm]
			\subfloat[(d)]
			{\includegraphics[width = \patchSize, height = \patchSize]
				{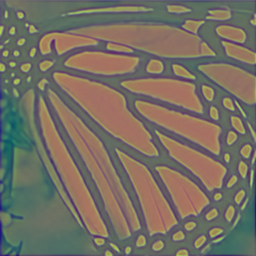}} &
			\subfloat[(e)]
			{\includegraphics[width = \patchSize, height = \patchSize]
				{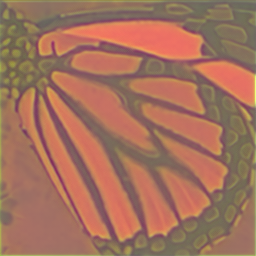}} &
			\subfloat[(f)]
			{\includegraphics[width = \patchSize, height = \patchSize]
				{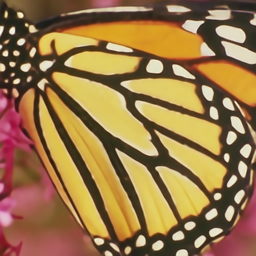}}
		\end{tabular}
	\end{center}
	\setlength{\abovecaptionskip}{0pt plus 2pt minus 2pt}
	\setlength{\belowcaptionskip}{0pt plus 2pt minus 2pt}
	\caption{Visualizations of different iterations with \textbf{BI}$\times4$ degradation. (a)-(e): $\mathbf{I}^{SR}_k$ of iteration $k=1,2,3,4,5$; (f): The final result $\mathbf{I}^{SR}$ after MLE.}
	\label{Fig:vis-norm}
\end{figure}

Moreover, the feature maps $\mathbf{I}^{SR}_k$ for different iteration $k$ are analyzed. The visualization results of each $\mathbf{I}^{SR}_k$ and output image $\mathbf{I}^{SR}$ are shown in Fig.~\ref{Fig:ablation-iter}. From the visual quality comparison, it could be found that with the increase of iteration, richer details can be found on the feature maps. In the first and second iterations, there are structural information with clear lines and contours. In the next three iterations, there are more details on the wing. It is also observed that different iterations concentrate on different features. This is in line with the hypothesis of descent direction in \textit{Solver LR} for each iteration. The MLE fuses results from every iteration and makes full use of these results. We train the network end-to-end and only consider the criterion between $\mathbf{I}^{SR}$ and $\mathbf{I}^{HR}$. There is no extra supervised constrains on different solvers. This is why Fig.~\ref{Fig:vis-norm} looks unnatural. In fact, $\{\mathbf{I}^{SR}_k\}^K_{k=1}$ span a solution space and the unnatural textures focus on specific directions for finding a suitable $\textbf{I}^{SR}$.

In this situation, the number of iterations is fixed. Different iterations find different descent directions, and \textit{Solver MLE} learns how to comprehensively consider the diverse directions. Since \textit{Solver MLE} is fixed, $K$ is fixed.

\begin{figure}
	\captionsetup[subfloat]{labelformat=empty, justification=centering}
	\begin{center}
		\scriptsize
		\setlength\tabcolsep{0.1cm}
		\begin{tabular}[b]{c@{\hspace{0.1cm}}c@{\hspace{0.1cm}}c}
			{\includegraphics[width=0.25\linewidth]
				{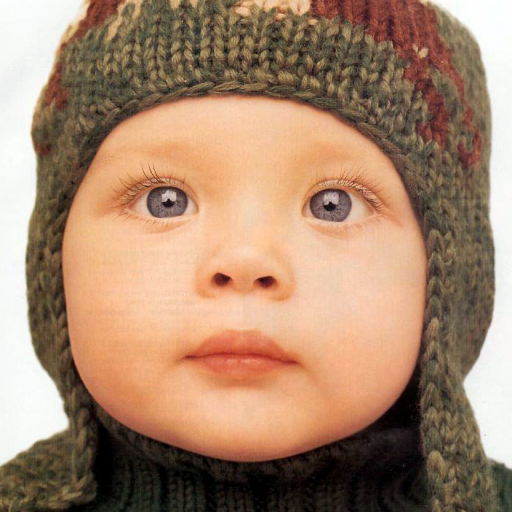}} &
			{\includegraphics[width=0.25\linewidth]
				{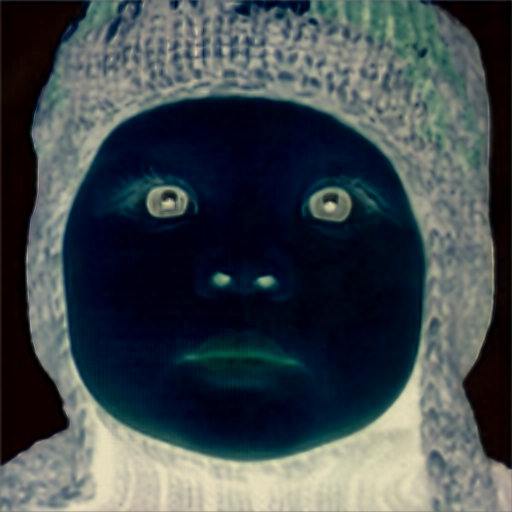}} &
			{\includegraphics[width=0.25\linewidth]
				{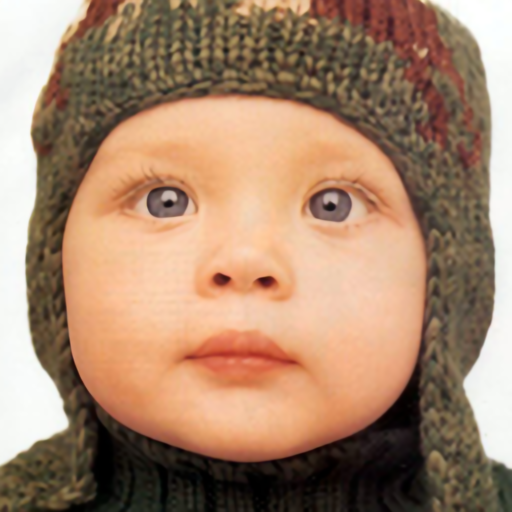}} \\
			
			{\includegraphics[width=0.25\linewidth]
				{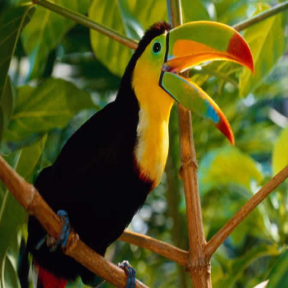}} &
			{\includegraphics[width=0.25\linewidth]
				{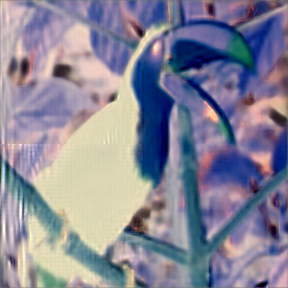}} &
			{\includegraphics[width=0.25\linewidth]
				{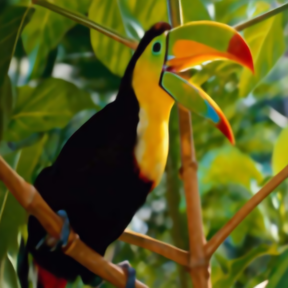}} \\[-0.35cm]
			
			\subfloat[(a)~HR image]{\includegraphics[width=0.25\linewidth]
				{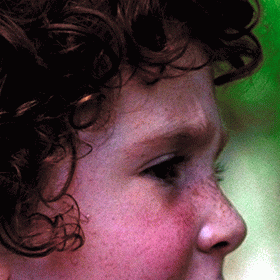}} &
			\subfloat[(b)~Plug-and-Play]{\includegraphics[width=0.25\linewidth]
				{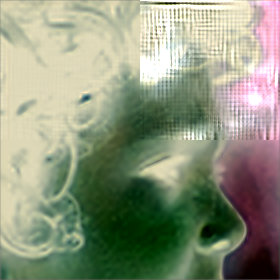}} &
			\subfloat[(c)~End-to-End]{\includegraphics[width=0.25\linewidth]
				{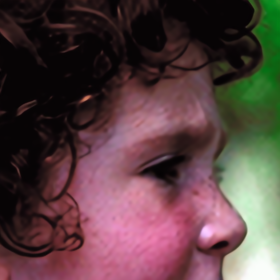}} 
		\end{tabular}
	\end{center}
	\setlength{\abovecaptionskip}{0pt plus 2pt minus 2pt}
	\setlength{\belowcaptionskip}{0pt plus 2pt minus 2pt}
	\caption{Visual quality comparisons of Plug-and-Play and End-to-End with \textbf{BI}$\times4$ degradation.}
	\label{Fig:vis-plug-and-play}
\end{figure}

\textbf{Study on Plug-and-Play}. To better illustrate the hypothesis about plug-and-play, \textit{Solver SR} is substituted with a pre-trained RCAN~\cite{rcan_zhang2018image}. The pre-trained model is downloaded from GitHub repository provided by the author. Notice that there is no change except for the \textit{Solver SR}. The result is shown in Fig.~\ref{Fig:vis-plug-and-play}. The images are chosen from Set5 dataset. From the results, plug-and-play cannot deliver satisfactory result on both color and texture, which provide useful evidence on the hypothesis of learning different coding directories.

\subsection{Limitation}
As discussed in the experiments, ISRN achieves competitive or better PSNR/SSIM results than recent methods, but the perceptual restoration of ISRN is no better than the GAN-based methods. In this point of view, it is challenging for ISRN to preserve more high-frequency information and fine details than advanced GAN-based methods. This is because this network is designed based on the optimization formulations, and we do not consider the GAN-loss functions while training.

\section{Conclusion}
In this paper, a novel iterative super-resolution network (ISRN) was proposed for SISR problem. We analyzed the problem from an optimization perspective, and found a feasible solution in an iterative manner. Based on the formulation study, each module of ISRN was elaborately designed, and a maximization likelihood estimation was performed to considerate results from all iterations. Specifically, a novel block named FNB with feature normalization was introduced to compose the network, and grouped in a residual-in-residual way. Considering the drawbacks of batch normalization, the feature normalization (FN) was designed for feature regulation with depth-wise convolution. Extensive experimental results on benchmark datasets with different degradation models show that the proposed ISRN and extension model ISRN$^+$ are able to recover structural information more effectively, and to achieve competitive or better performance with much fewer parameters.

Our future work will try to extend ISRN to a GAN-based method with specific discriminator and losses. Specially, detail enhancement components will be considered to preserve more high-frequency information and fine details.


%





\ifCLASSOPTIONcaptionsoff
  \newpage
\fi



\bibliographystyle{IEEEtran}
\bibliography{IEEEabrv,egbib}
%



%

\vspace{-10mm}
\begin{IEEEbiography}[{\includegraphics[width=1in,height=1.25in,clip,keepaspectratio]{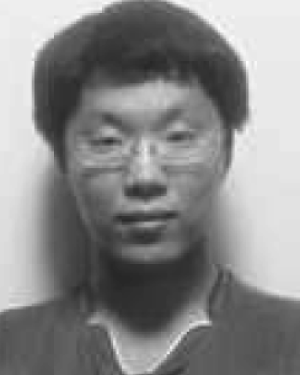}}]{Yuqing Liu}
received the B.S. degree in software engineering from the Dalian University of Technology, China, in 2017. He is currently pursuing the Ph.D. degree. His current research interests include video compression, processing, and analysis.
\end{IEEEbiography}

\begin{IEEEbiography}[{\includegraphics[width=1in,height=1.25in,clip,keepaspectratio]{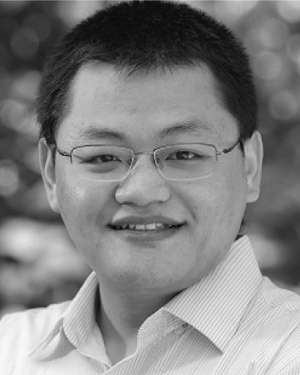}}]{Shiqi Wang}
(M’15) received the B.S. degree in computer science from the Harbin Institute of Technology in 2008, and the Ph.D. degree in computer application technology from the Peking University, in 2014. From 2014 to 2016, he was a Post-doctoral Fellow with the Department of Electrical and Computer Engineering, University of Waterloo, Waterloo, Canada. From 2016 to 2017, he was with the Rapid-Rich Object Search Laboratory, Nanyang Technological University, Singapore, as a Research Fellow. He is currently an Assistant Professor with the Department of Computer Science, City University of Hong Kong. He has proposed over 40 technical proposals to ISO/MPEG, ITU-T, and AVS standards. His research interests include video compression, image/video quality assessment, and image/video search and analysis.
\end{IEEEbiography}

\begin{IEEEbiography}[{\includegraphics[width=1in,height=1.25in,clip,keepaspectratio]{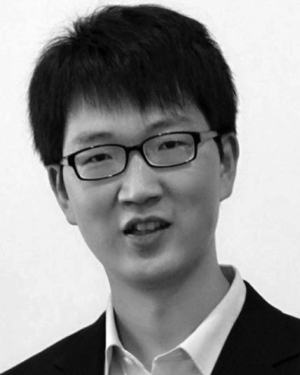}}]{Jian Zhang} (M'14) received the B.S. degree from the Department of Mathematics, Harbin Institute of Technology (HIT), Harbin, China, in 2007, and received his M.Eng. and Ph.D. degrees from the School of Computer Science and Technology, HIT, in 2009 and 2014, respectively. 

Currently, he is an Assistant Professor with the School of Electronic and Computer Engineering, Peking University Shenzhen Graduate School, Shenzhen, China. His research interests include intelligent multimedia processing, deep learning and optimization. He received the Best Paper Award at the 2011 IEEE Visual Communications and Image Processing (VCIP) and was a co-recipient of the Best Paper Award of 2018 IEEE MultiMedia.
\end{IEEEbiography}

\begin{IEEEbiography}[{\includegraphics[width=1in,height=1.25in,clip,keepaspectratio]{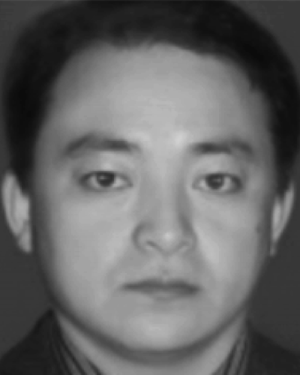}}]{Shanshe Wang}
 received the B.S. degree from the Department of Mathematics, Heilongjiang University, Harbin, China, in 2004, the M.S. degree in computer software and theory from Northeast Petroleum University, Daqing, China, in 2010, and the Ph.D. degree in computer science from the Harbin Institute of Technology. He held a postdoctoral position at Peking University, Beijing, from 2016 to 2018. He joined the School of Electronics Engineering and Computer Science, Institute of Digital Media, Peking University, where he is currently a Research Associate Professor. His current research interests include video compression and image and video quality assessment.
\end{IEEEbiography}

\begin{IEEEbiography}[{\includegraphics[width=1in,height=1.25in,clip,keepaspectratio]{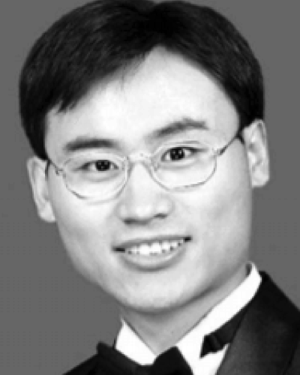}}]{Siwei Ma}
(S’03–M’12) received the B.S. degree from Shandong Normal University, Jinan, China, in 1999, and the Ph.D. degree in computer science from the Institute of Computing Technology, Chinese Academy of Sciences, Beijing, China, in 2005. From 2005 to 2007, he held a post-doctoral position at the University of Southern California, Los Angeles, USA. Then, he joined the Institute of Digital Media, School of Electronics Engineering and Computer Science, Peking University, Beijing, where he is currently a Professor. He has published over 100 technical articles in refereed journals and proceedings in the areas of image and video coding, video processing, video streaming, and transmission.
\end{IEEEbiography}

\begin{IEEEbiography}[{\includegraphics[width=1in,height=1.25in,clip,keepaspectratio]{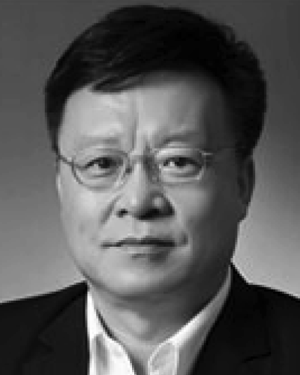}}]{Wen Gao}
 (M’92–SM’05–F’09) received the Ph.D. degree in electronics engineering from The University of Tokyo, Japan, in 1991. He was a Professor of computer science with the Harbin Institute of Technology, from 1991 to 1995, and a Professor with the Institute of Computing Technology, Chinese Academy of Sciences. He is currently a Professor of computer science with Peking University, China.. He has published extensively including five books and over 600 technical articles in refereed journals and conference proceedings in the areas of image processing, video coding and communication, pattern recognition, multimedia information retrieval, multimodal interface, and bioinformatics. He chaired a number of prestigious international conferences on multimedia and video signal processing, such as the IEEE ICME and the ACM Multimedia, and also served on the advisory and technical committees of numerous professional organizations. He served or serves on the Editorial Board for several journals, such as the IEEE Transactions on Circuits and Systems for Video Technology, the IEEE Transactions on Multimedia, the IEEE Transactions on Image Processing, the IEEE Transactions on Autonomous Mental Development, the EURASIP Journal of Image Communications, and the Journal of Visual Communication and Image Representation.
\end{IEEEbiography}







\end{document}